\documentclass[lettersize,journal]{IEEEtran}
\IEEEoverridecommandlockouts

\usepackage{amsmath,amssymb,amsfonts}
\usepackage{cite}
\usepackage{algorithm}
\usepackage{algorithmicx}
\usepackage{algpseudocode}
\usepackage{stfloats}
\usepackage{graphicx}
\usepackage{textcomp}
\usepackage{xcolor}
\usepackage{amsthm}
\usepackage{amsmath}
\usepackage{booktabs}
\usepackage{hyperref}
\usepackage{cases}
\usepackage{comment}
\usepackage{empheq}
\usepackage{caption}
\usepackage{subcaption}
\usepackage{cuted}
\usepackage{url}

\allowdisplaybreaks[4]

\usepackage{mathtools}

\allowdisplaybreaks[4]

\def\BibTeX{{\rm B\kern-.05em{\sc i\kern-.025em b}\kern-.08em
    T\kern-.1667em\lower.7ex\hbox{E}\kern-.125emX}}
\begin{document}

\title{Movable Antenna Enhanced Federated Fine-Tuning of Large Language Models via Hybrid Client Selection Optimization}

\author{Yang Zhao\textsuperscript{1},~\IEEEauthorblockN{Yue Xiu\textsuperscript{2},~\textup{Yue Xiu}$^{1}$,~\textup{Dusit Niyato}$^{1}$}
\IEEEauthorblockA{\textsuperscript{1}College of Computing and Data Science,
Nanyang Technological University, Singapore\\
\textsuperscript{2}University of Electronic Science and Technology of China, Chengdu, China \\
\textsuperscript{3}Lassonde School of Engineering, York University, Canada
}
}

\author{
\IEEEauthorblockN{Yang Zhao,~\IEEEmembership{Member,~IEEE}, Yue~Xiu,~\IEEEmembership{Member,~IEEE}, Chengxiao Dai,~\IEEEmembership{Student Member,~IEEE}, Dusit~Niyato,~\IEEEmembership{Fellow,~IEEE}, Ning Wei,~\IEEEmembership{Member,~IEEE}
}
\thanks{Yang Zhao.(e-mail:zhao0466@e.ntu.edu.sg).}
\thanks{Yue Xiu and Ning Wei are with the National Key
Laboratory of Science and Technology on Communications, University of
Electronic Science and Technology of China, Chengdu 611731, China (e-mail:
xiuyue12345678@163.com).}
\thanks {Chengxiao Dai is with the University of Sydney, Australia. (e-mail:cdai0023@uni.sydney.edu.au)}
\thanks{Dusit Niyato is with Nanyang Technological University, Singapore.
 (e-mail:dniyato@ntu.edu.sg).} 
}

\maketitle

\begin{abstract}
Federated fine-tuning of large language models (LLMs) over bandwidth-limited 6G links must meet strict round-time and energy budgets. Analog over-the-air (OTA) aggregation reduces uplink cost but is sensitive to fading and interference, which distort the aggregated gradient. We consider a two-phase workflow, centralized pre-training followed by federated fine-tuning, where the base station uses a movable antenna (MA) array. In each round, MA element positions and the receive/transmit beamformers are adjusted under minimum-spacing constraints to reshape the channel and improve OTA aggregation without increasing user transmit power. We formulate a mixed-integer, nonconvex resource-allocation problem that jointly selects clients and optimizes the numbers of global rounds, CPU frequencies, mini-batch sizes, MA positions, and analog weights under end-to-end latency and energy limits. A successive convex approximation–penalty dual decomposition (SCA–PDD) routine alternates convex updates with oblique-manifold beamforming and spacing-aware MA placement. Experiments on OpenLLaMA\mbox{-}v2 (3B) with LoRA and 4-bit quantization on Alpaca and Dolly (10 clients) attain round-30 validation perplexities as low as \textbf{2.94} (Alpaca, $K{=}1$) and \textbf{4.62} (Dolly, $K{=}1$). Relative to the strongest non-MA baseline at the same concurrency, this corresponds to \textbf{17.4\%} (Alpaca, $K{=}1$) and \textbf{54.4\%} (Dolly, $K{=}1$) lower perplexity; at $K{=}2$ the reductions are \textbf{14.2\%} (Alpaca) and \textbf{13.7\%} (Dolly). Participation fairness also improves across all uplink concurrencies $K\in\{1,2,4,8\}$ (where $K$ is the number of clients transmitting concurrently per OTA round), with the largest margins when fewer clients transmit per round.
\end{abstract}


\section{Introduction}

Large language models (LLMs) based on GPT-like architectures achieve strong performance on generation, question answering, and summarization~\cite{matarazzo2025surveylargelanguagemodels}. They are typically trained in two stages: a phase of \emph{pre-training} on massive, general-purpose corpora~\cite{gao2020pile800gbdatasetdiverse}, followed by \emph{fine-tuning} on task- or domain-specific data, for example, biomedical, legal, or medical~\cite{upadhyay2025synlexlmscalinglegalllms,lee2025medllm}. In wireless edge settings targeted by 6G, fine-tuning must respect tight round-time and energy budgets, making the joint management of communication and computation a central challenge.

A common approach is to split training into (i) centralized pre-training on a compute-rich server and (ii) federated fine-tuning across many devices~\cite{ghiasvand2025decentralizedlowrankfinetuninglarge,lin2024splitlorasplitparameterefficientfinetuning}. Although this two-phase design limits data movement, iterative model synchronization remains a bottleneck under fading and interference. Analog over-the-air (OTA) gradient aggregation reduces uplink bandwidth by superposing waveforms, but it introduces two physical-layer distortions: (i) an \emph{analog-sum mismatch} between the received superposition and the desired weighted gradient sum, and (ii) \emph{post-combining noise}. Both degrade the effective gradient used by SGD and can slow or destabilize learning unless the wireless configuration is adapted round by round.

To mitigate these distortions, we consider a base station equipped with a movable antenna (MA) array. Unlike fixed phased arrays or purely reflective surfaces, MAs physically reposition radiating elements within a constrained region. By adjusting element positions and beamformers slightly each round, the base station can reshape the channel to better align simultaneous uplink transmissions, thereby reducing analog-sum mismatch and shaping post-combining noise without increasing user transmit power. Realizing these gains requires coordinating MA geometry with higher-layer training choices under practical constraints such as minimum inter-element spacing, limited displacement per round, motion latency/actuator energy, and channel-estimation overhead.

Another key factor is the distribution shift: the domain-specific fine-tuning data often differ statistically from the pre-training corpus~\cite{dorfner2025biomedical,ling2024domainspecializationkeymake}. We use a Wasserstein-based term to quantify this shift and incorporate it in a convergence view that clarifies how pre-training progress, OTA-induced distortions, and domain mismatch jointly affect the final loss.

Classical wireless federated learning largely optimizes communication rounds, device scheduling, and power control under fixed array geometries, without geometry‑aware channel reshaping at the base station \cite{dinh2021flwn,nishio2019fedcs,liu2021unified,yang2022fl6g}. OTA aggregation papers commonly adopt simplified channels (LoS or i.i.d. Rayleigh) and static arrays; analyses typically model additive Gaussian noise and do not couple it to array geometry or motion, and most works omit motion, solver, or CSI‑estimation overheads from the end‑to‑end budget \cite{yang2020airfl,xiao2024oafl,zeng2019aircomp,li2019aircompmimo}. Meanwhile, MA studies demonstrate that repositioning elements can reshape channels, but integration with FL/OTA objectives remains limited \cite{mei2024magraph,wei2025mairs,zhu2024maenhanced}. Finally, recent efforts on federated fine‑tuning of LLMs focus on privacy and communication aspects and do not jointly design OTA aggregation with MA geometry under tight latency/energy and distribution shift \cite{zhang2024fedgpt,yao2024fedllm}.

\textbf{Contributions.} Our main contributions are as follows.
\begin{itemize}
  \item We equip the server with a movable-antenna array and reconfigure element positions and receive/transmit beamformers each round to reduce analog-sum mismatch and shape post-combining noise under minimum-spacing constraints.
  \item We jointly optimize the numbers of global rounds, client participation, CPU frequencies, mini-batch sizes, MA positions, and analog weights under latency and energy limits.
  \item We present a bound in which OTA noise enters explicitly, and a Wasserstein term accounts for the shift from pre-training to on-device data.
  \item A successive convex approximation–penalty dual decomposition (SCA–PDD) scheme updates continuous variables with convex surrogates, refines beamformers on an oblique manifold, and repositions MA elements with spacing-aware updates.
  \item Under a line-of-sight channel model, we benchmark against fixed-array OTA, digital FedAvg/OFDMA, Top-$K$ SNR, Gibbs sampling, and an MA-greedy heuristic, and report model quality and participation fairness in OpenLLaMA‑v2 (3B) with LoRA and 4-bit quantization over Alpaca and Dolly.
\end{itemize}

The remainder of this paper is organized as follows. Section~\ref{sec:related} reviews the state of the art in MA technology, federated fine-tuning of LLMs, and OTA aggregation. Section~\ref{sec:system_model} details the overall system model, including the two-phase training workflow, the MA architecture, and the associated latency and energy framework. Section~\ref{sec:problem} formulates a mixed integer non-convex resource allocation problem that couples learning, communication, and antenna geometry decisions. Section~\ref{sec:Algorithm} describes a hybrid SCA-PDD to solve this problem. Section~\ref{sec:num-results} presents numerical results that validate the proposed framework, and Section~\ref{sec:conclusion} concludes the article. Table~\ref{notations} summarizes all notation.

\begin{table}[h!]
\caption{Notation Table.}
\begin{tabular}{{|l|l|}}
\hline
\textbf{Symbol} & \textbf{Description} \\
\hline
$\alpha^2, \hat{\alpha}^2$ & Gradient-variance bounds (pre-/fine-tuning). \\\hline
$\beta_{u,t}$ & Complex path gain for user $u$ at time $t$. \\\hline
$\boldsymbol{\varepsilon}^{(n)}$ & OTA noise in round $n$. \\\hline
$C$ & FLOPs per sample. \\\hline
$d_w$ & Dimension of model parameters. \\\hline
$f_{u}^{(n)}$ & CPU frequency of user $u$ in round $n$. \\\hline
$g_{u,t}$ & Effective channel gain for user $u$. \\\hline
$\gamma,\hat{\gamma}$ & Step sizes (pre-/fine-tuning). \\\hline
$i,j$ & Indices of movable-antenna elements. \\\hline
$\kappa_0$ & CPU power coefficient. \\\hline
$L_{\mathrm{pre}}, L_{\mathrm{fine}}$ & Population losses (pre-training, fine-tuning). \\\hline
$\lambda$ & Carrier wavelength. \\\hline
$m \in \{0,\dots,M-1\}$ & Index for pre-training rounds (Phase I). \\\hline
$M \in \mathbb{Z}_+$ & Number of pre-training rounds. \\\hline
$n \in \mathcal{N}=\{0,\dots,N-1\}$ & Index for fine-tuning rounds (Phase II). \\\hline
$N \in \mathbb{Z}_+$ & Number of fine-tuning rounds. \\\hline
$N_T$ & Number of movable-antenna elements. \\\hline
$\nabla \tilde{L}_{\mathrm{pre}}, \nabla \tilde{L}_{\mathrm{fine}}$ & Stochastic gradients in each phase. \\\hline
$P_a$ & Max transmit power per user. \\\hline
$\rho, \hat{\rho}$ & Smoothness constants (pre-/fine-tuning). \\\hline
$\rho_{\mathrm{dist}}$ & Lipschitz constant w.r.t.\ data domain. \\\hline
$\sigma^2$ & Noise variance for OTA aggregation. \\\hline
$t$ & Time index in the channel model. \\\hline
$u \in \mathcal{U}=\{1,\dots,U\}$ & Index of distributed users/nodes. \\\hline
$U$ & Total number of distributed users. \\\hline
$\mathbf{a}(\cdot)$ & Array response function. \\\hline
$\mathbf{g}_{u,t}$ & Local gradient from user $u$ at time $t$. \\\hline
$\mathbf{h}_{u,t}$ & Channel vector from user $u$ at time $t$. \\\hline
$\mathbf{q}_t$ & Beamforming vector at the server (norm-1). \\\hline
$\mathbf{w}^{(m)}$ & Model parameters after $m$-th pre-training round. \\\hline
$\mathbf{w}^{(M+n)}$ & Model parameters after $n$ fine-tuning rounds. \\\hline
$\mathbf{x}_{n,i}$ & 2D location of $i$-th antenna in round $n$. \\\hline
$\mathbf{z}_A, \mathbf{z}_B$ & Continuous vs.\ discrete variables in SCA--PDD. \\\hline
$\Delta_{\mathrm{pre}}, \Delta_{\mathrm{fine}}$ & Variance-related overhead in each phase. \\\hline
$\eta_t$ & Reception-scaling factor. \\\hline
$\mathcal{E}(\cdot)$ & Total time and energy consumption functions. \\\hline
$\mathcal{L}(\cdot)$ & Total time and energy consumption functions. \\\hline
$\mathcal{L}_{\max}, Q_{\max}$ & Total time and energy limits. \\\hline
$\mathcal{U}_n \subseteq \{1,\dots,U\}$ & User set in fine-tuning round $n$. \\\hline
$W^{(m)}, b_{u}^{(n)}$ & Mini-batch size. \\\hline
$\mathrm{W}(P_{\mathrm{pre}},P_{\mathrm{fine}})$ & Wasserstein distance for distribution shift. \\
\hline
\end{tabular}
\label{notations}
\end{table}

\section{Related Work}
\label{sec:related}

In this section, we present a review of recent advances in movable antenna technology, communication-efficient federated fine-tuning of large language models, and OTA gradient aggregation, to establish the foundation for our proposed framework.

\subsection{Movable Antenna Technology}

Recent advances in reconfigurable antenna architectures have led to the concept of MAs, which allow the physical repositioning of antenna elements to adapt to channel variations~\cite{10962171,zhu2025tutorialmovableantennaswireless}. Unlike traditional phased arrays, MAs offer a new degree of freedom in geometry, allowing dynamic adaptation to environmental changes such as user movement, blockage, or interference. By continuously adjusting the antenna position, orientation or structure within a predefined surface or volume, MAs can achieve highly flexible spatial configurations that enhance the robustness of the link in non-stationary~\cite{10684758}.

Previous research indicates that even millimeter-level adjustments in antenna position can significantly enhance line-of-sight (LoS) connectivity, reduce multipath fading, and lower interference in dense or dynamic wireless environments~\cite{10962171,wang2024movableantennaschannelmeasurement}. These gains are particularly relevant in high-frequency systems, where propagation conditions change rapidly and coverage is highly sensitive to geometry.

Moreover, the integration of MAs with intelligent reflecting surfaces (IRSs) has been explored as a means of further enhancing spatial diversity and signal delivery~\cite{liu2023reconfigurable}. Such combinations can maximize coverage, improve spatial multiplexing capabilities, and mitigate signal blockage in complex wireless environments. Compared to traditional designs that treat antenna positioning and beamforming as decoupled problems~\cite{liu2023reconfigurable} joint optimization strategies that use MA mobility have demonstrated improved communication reliability and spectrum efficiency.

Together, MAs represent a paradigm shift from static array configurations to geometry-aware, environment-adaptive architectures. This makes them a promising foundation for next-generation wireless systems, particularly in applications such as federated learning or OTA aggregation, where spatial robustness and transmission precision are critical~\cite{zhao2025movableantennaaidedfederatedlearning}.

\subsection{Federated Fine-Tuning of Large Language Models}

Although LLMs have shown impressive capabilities in generating fluent and contextually coherent natural language text across a wide range of tasks, effectively deploying these models in real-world scenarios often requires fine-tuning tailored to specific tasks or domains~\cite{Devlin2019BERT,Kopf2021TwoPhaseLLM}. This requirement is particularly pronounced in vertical applications such as healthcare, law, or engineering, where vocabulary, syntax, and knowledge distributions differ substantially from those found in general-purpose corpora, often resulting in degraded performance without domain-specific adaptation.

Then, FL has emerged as a promising paradigm for addressing data privacy and governance challenges, especially in scenarios involving sensitive or proprietary data~\cite{Yang2019Federated,yu2022differentiallyprivatefinetuninglanguage,10.5555/3600270.3601038}. In such settings, data remain on the local device and only model updates are communicated, which preserves privacy and ensures compliance with locality-aware data regulations.

However, iterative model synchronization introduces high communication overhead, which poses a significant bottleneck, particularly for LLMs that comprise billions of parameters~\cite{Alnaasan2024Communication}. Limited uplink bandwidth, heterogeneous device capabilities, and unreliable wireless channels further exacerbate the problem, potentially causing slow convergence, degraded model quality, or even training divergence~\cite{zhang2024buildingfederatedgptfederated}. These challenges underscore the urgent need for scalable, communication-efficient, and resource-adaptive approaches to the federated fine-tuning of LLMs.

\subsection{Over-the-Air Gradient Aggregation}

OTA aggregation exploits waveform superposition in wireless channels to merge gradients from multiple devices in a single uplink resource block~\cite{Zhu2020OTAsurvey}. This technique dramatically reduces transmission overhead in FL but is vulnerable to fading and interference. Existing work addresses these issues through power control and beamforming~\cite{Chen2021BeamformingOTA}, yet few studies have examined the use of MAs to enhance OTA performance.

In contrast to the existing literature that addresses MAs, federated LLM fine-tuning, and OTA aggregation separately, our work unifies these three directions. Specifically, we show how the geometries of MAs can be jointly optimized with model training schedules and OTA updates, leading to improved convergence under stringent time and energy budgets.

\section{System Model}
\label{sec:system_model}

\begin{figure*}
    \centering
    \includegraphics[width=\linewidth]{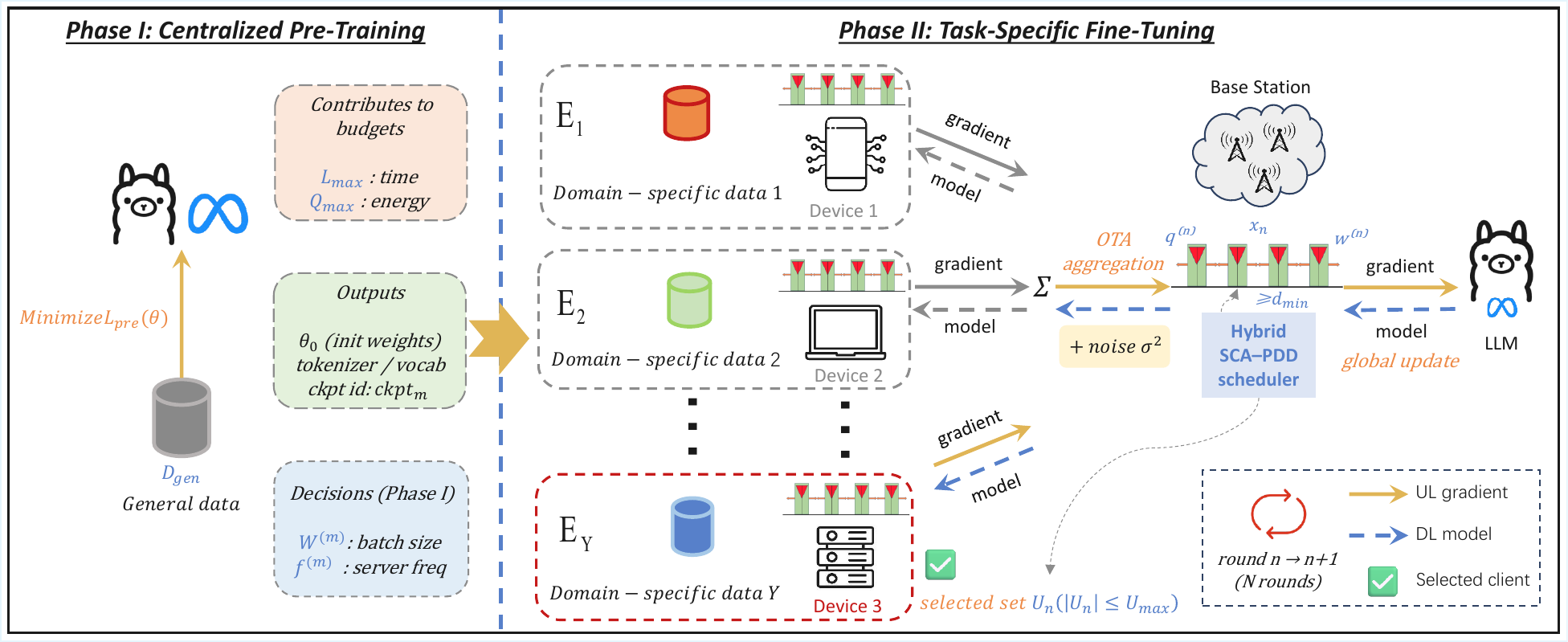}
    \caption{Movable antenna enhanced federated fine-tuning framework for LLMs. Phase I (left) shows centralized pre-training in which an edge server trains a foundation LLM on a large, general-purpose dataset. Phase II (right) depicts task-specific fine-tuning.}
    \label{fig:enter-label}
\end{figure*}

We consider a two-phase learning workflow that blends (1) centralized pre-training of an LLM on an edge server with ample compute and storage, and (2) wireless federated fine-tuning on $U$ clients that hold domain-specific data but face bandwidth and energy constraints  as shown in Figure~\ref{fig:enter-label}. OTA gradient aggregation is used in the second phase, and the server is equipped with a geometry--reconfigurable  MA array that improves signal superposition without increasing transmit power~\cite{8952884}.

\subsection{Phase I: Centralized Pre-Training}
\label{subsec:pretrain}

Let $\mathbf{w}^{(m)} \in \mathbb{R}^{d_w}$ denote the model parameters after the $m$-th pre-training round, where $m \in \{0,\dots,M-1\}$ and $d_w$ is the dimension of the parameter vector. Define $\mathcal{D}_{\mathrm{pre}}$ as the server's dataset, which consists of samples drawn from a distribution $D$. The population loss for pre-training is then given by
\begin{equation}
L_{\mathrm{pre}}\bigl(\mathbf{w}^{(m)}\bigr)
\;=\;
\mathbb{E}_{d \sim D}\Bigl[\ell\bigl(\mathbf{w}^{(m)}, d\bigr)\Bigr].
\label{eq:pop_loss_pre}
\end{equation}

In practice, an empirical approximation of \eqref{eq:pop_loss_pre} is used via a mini-batch $\mathcal{B}^{(m)} \subseteq \mathcal{D}_{\mathrm{pre}}$ of size $W^{(m)}$. Denoting the empirical loss over this mini-batch by
\begin{equation}
\tilde{L}_{\mathrm{pre}}\!\bigl(\mathbf{w}^{(m)}, \mathcal{B}^{(m)}\bigr)
\;=\;
\frac{1}{W^{(m)}} \sum_{d \in \mathcal{B}^{(m)}} \ell\bigl(\mathbf{w}^{(m)}, d\bigr).
\label{eq:emp_loss_pre}
\end{equation}
Subsequently, the model is updated via a mini-batch stochastic gradient descent (SGD) step as follows:
\begin{equation}
\mathbf{w}^{(m+1)} 
\;=\; 
\mathbf{w}^{(m)}
\;-\; 
\gamma \,\nabla_{\mathbf{w}^{(m)}} \,\tilde{L}_{\mathrm{pre}}\!\bigl(\mathbf{w}^{(m)}, \mathcal{B}^{(m)}\bigr),
\label{eq:update_rule_pre}
\end{equation}
where $\gamma$ is the (constant) step size for Phase~I and $\nabla \tilde{L}_{\mathrm{pre}}$ denotes the stochastic gradient with respect to $\mathbf{w}^{(m)}$. After $M$ such rounds, the final pre-trained model is
\begin{align}
\mathbf{w}^{(M)}
\;\equiv\;
\mathbf{w}^{\mathrm{pre\text{-}train}}.
\label{eq:final_pretrain}
\end{align}

\subsection{Phase II: Task-Specific Fine-Tuning}
\label{subsec:fine}

After centralized pre-training, the model is refined over $N$ fine-tuning rounds, each indexed by $n \in \mathcal{N}$. Let $\mathbf{w}^{(M+n)}$ denote the parameters in round~$n$. Suppose $U$ distributed users participate in the fine-tuning process, each holding data drawn from a possibly shifted distribution. For a given user $u \in \mathcal{U}$, define
\begin{equation}
L_{u}\bigl(\mathbf{w}^{(M+n)}, \mathcal{B}_{u}^{(n)}\bigr)
\;=\;
\mathbb{E}_{d \,\sim\, \hat{\mathcal{P}}_{u}}
\Bigl[\ell\bigl(\mathbf{w}^{(M+n)}, d\bigr)\Bigr],
\end{equation}
where $\mathcal{B}_{u}^{(n)}$ is the mini-batch for user~$u$ in the round~$n$, and $\hat{\mathcal{P}}_{u}$ denotes the local data distribution. An aggregate loss over all users is then
\begin{equation}
L_{\mathrm{fine}}\bigl(\mathbf{w}^{(M+n)}\bigr)
\;=\;
\frac{1}{U}\,\sum_{u=1}^{U}\,
L_{u}\bigl(\mathbf{w}^{(M+n)}, \mathcal{B}_{u}^{(n)}\bigr),
\end{equation}
where $\bigcup_{u=1}^{U} \mathcal{B}_{u}^{(n)}$ comprises the overall mini-batch in round~$n$.  

The stochastic gradient of the local loss for user~$u$ is computed by summing the sample-wise derivatives over $\mathcal{B}_{u}^{(n)}$ (of size $b_{u}^{(n)}$). Specifically,
\begin{equation}
\begin{split}
\nabla \tilde{L}_{u}^{(n)}\bigl(\mathbf{w}^{(M+n)}, \mathcal{B}_{u}^{(n)}\bigr)
&= \frac{1}{b_{u}^{(n)}} 
\sum_{d_i \in \mathcal{B}_{u}^{(n)}} \\
&\quad \times\,
\nabla_{\mathbf{w}^{(M+n)}} \,\ell\!\bigl(\mathbf{w}^{(M+n)}, d_i\bigr).
\end{split}
\end{equation}
Then, the fine-tuning gradient can be formed by weighting each user’s gradient according to its mini-batch size :
\begin{equation}
\begin{split}
\nabla L_{\mathrm{fine}}\!\bigl(\mathbf{w}^{(M+n)}, \{\mathcal{B}_{u}^{(n)}\}\bigr)
&= \sum_{u=1}^{U}
\frac{b_{u}^{(n)}}{\sum_{v=1}^{U} b_{v}^{(n)}} \\
&\quad \times\,
\nabla L_{u}^{(n)}\!\bigl(\mathbf{w}^{(M+n)}, \mathcal{B}_{u}^{(n)}\bigr),
\end{split}
\end{equation}
where $\nabla L_{u}^{(n)}$ is the empirical gradient for user~$u$ in round~$n$. Finally, the parameters are updated by
\begin{equation}
\mathbf{w}^{(M+n+1)}
\;=\;
\mathbf{w}^{(M+n)}
\;-\;
\hat{\gamma}\,\nabla L_{\mathrm{fine}}
\!\bigl(\mathbf{w}^{(M+n)}, \{\mathcal{B}_{u}^{(n)}\}\bigr),
\end{equation}
where $\hat{\gamma}$ is the step size in Phase~II.  At the end of $N$ such rounds, one obtains $\mathbf{w}^{(M+N)}$ as the fine-tuned model specific to the task.

\subsection{Communication Model}\label{subsec:comm}
In the communication part of our system, we optimize the position of the MA element within the container to maximize channel gain or improve the signal-to-noise ratio under varying channel conditions. Additionally, we design a selection strategy to determine the optimal antenna position at each transmission instance to enhance communication reliability, and we also jointly optimize the beamforming vector in coordination with the MA's location.

The uplink transmission rate for user~$u$ in round~$n$ is then given by
\begin{equation}
R_{ul}^{(n),u}=\;
B_{ul}^{u}\,\log_{2}\!\Bigl(
1 \;+\;
\frac{(\boldsymbol{q}^{(n)})^{H}\boldsymbol{h}_{ul}^{u}(\boldsymbol{x}^{(n)})\,p_{u}^{(n)}}{B_{ul}^{u}\,N_{ul}^{n}}
\Bigr),
\label{eq:ul_rate}
\end{equation}
where $\boldsymbol{q}^{(n)}$ is the combine beamforming in the BS. $\boldsymbol{h}_{ul}^{u}(\boldsymbol{x}^{(n)})$ denote the uplink channel gain for user~$u$ in round~$n$. Based on the field response theory \cite{10318061}, the uplink light-of-sight (Los) channel $\boldsymbol{h}_{ul}^{u}(\boldsymbol{x}^{(n)})$ is modeled as
\begin{align}
&\boldsymbol{h}_{ul}^{u}(\boldsymbol{x}^{(n)})=\alpha_{u}^{(n),l}\boldsymbol{a}_{u}(\boldsymbol{x}^{(n)}).
\end{align}
The parameter $\alpha_{u}^{(n),l}$
represents the gain in the path of the uplink channel. The vector $\boldsymbol{x}^{(n)}=[x_{1}^{(n)},\cdots,x_{N_{t}}^{(n)}]^{T}$ denotes the MA position of the
BS, where $N_{t}$ is the number of MA for the BS. The vector $\boldsymbol{a}_{u}(\boldsymbol{x}^{(n)})$ is the response vector for the uplink channel array. According to the movable antenna model, the array response vector is expressed as
\begin{align}
&\boldsymbol{a}_{u}(\boldsymbol{x}^{(n)})=[e^{j\frac{2\pi}{\lambda}x_{1}^{(n)}\cos(\phi_{u})},\cdots,e^{j\frac{2\pi}{\lambda}x_{N_{t}}^{(n)}\cos(\phi_{u})}]^{T},
\end{align}
where $\phi_{u}$ is the angle of arrival (AoA) of the $u$-th user. $B_{ul}^{u}$ is the available uplink bandwidth and $N_{ul}^{(n)}$ is the noise spectral density. The downlink rate can be expressed as
\begin{equation}
R_{u}^{(n), dl}
\;=\;
B_{dl}\,\log_{2}\!\Bigl(
1 \;+\;
\frac{\boldsymbol{h}_{dl,u}^{H}(\boldsymbol{x}^{(n)})\,\boldsymbol{w}^{(n)}\tilde{P}}{B_{dl}\,\tilde{N}_{dl}^{(n)}}
\Bigr),
\label{eq:dl_rate}
\end{equation}
in which $B_{dl}$ is the bandwidth and $\tilde{P}$, $\tilde{N}_{dl}^{(n)}$ denote the server’s transmit power and downlink noise density for user~$u$ in round~$n$, respectively. Similarly, based on the field response theory \cite{wang2024movableantennaschannelmeasurement}, the downlink LoS channel $\boldsymbol{h}_{dl,u}(\boldsymbol{x}^{(n)})$ is
\begin{align}
&\boldsymbol{h}_{dl,u}(\boldsymbol{x}^{(n)})=\beta_{u}^{(n)}\boldsymbol{a}_{u}(\boldsymbol{x}^{(n)}).
\end{align}
The term $\beta_{u}^{(n)}$ represents the gains in the uplink and downlink channels. The vector $\boldsymbol{a}_{u}(\boldsymbol{x}^{(n)})$ denotes the response vector of the downlink channel array. Based on the MA model presented in \cite{10318061}, the array response vector is expressed as
\begin{align}
&\boldsymbol{a}_{u}(\boldsymbol{x}^{(n)})=[e^{j\frac{2\pi}{\lambda}x_{1}^{(n)}\sin(\theta_{u})},\cdots,e^{j\frac{2\pi}{\lambda}x_{N_{t}}^{(n)}\sin(\theta_{u})}]^{T},
\end{align}
where $\theta_{u}$ is the departure of angle (AoD) from the BS to the $u$ user. Both \eqref{eq:ul_rate} and \eqref{eq:dl_rate} follow standard Shannon-capacity formulas under Gaussian noise. In each round, the superposition of uplink transmissions underlies the aggregation of the OTA gradient. Beamforming or MA configurations can be used at the server to improve signal-to-noise ratios.

Let $u \in \mathcal{U}$ and $n \in \mathcal{N}$ index users and fine-tuning rounds, respectively. Let $p_{u}^{(n)}$ denote the transmit power of user~$u$ during the round~$n$. The average and instantaneous power constraints for each user become
\begin{equation}
\frac{1}{N}\,\sum_{n=0}^{N-1} p_{u}^{(n)}
\;\le\;
p_{u}^{\mathrm{ave}},
\quad
\forall\,u \in\mathcal{U},
\end{equation}
\begin{equation}
0 \;\le\; p_{u}^{(n)} \;\le\; P_{a},
\quad 
\forall\,n \in \mathcal{N},
\quad
\forall\,u \in \mathcal{U},
\end{equation}
where $p_{u}^{\mathrm{ave}}$ is the average transmit power budget and $P_{a}$ is the maximum transmit power allowed per user.

\subsection{Training Latency}\label{subsec:latency}

The end-to-end training delay equals the sum of two components: (i) server-side pre-training latency, the purely computational time for the edge server to process $M$ mini-batches of size $W^{(m)}$ at frequencies ${f^{(m)}}$; and (ii) client-side fine-tuning latency, dominated each round by the slowest user, which combines downlink reception of the global model, local SGD on a batch of size $b_{u}^{(n)}$ at frequency $f_{u}^{(n)}$, and uplink transmission of the resulting gradient. Specifically, during Phase~I, the total pre-training latency is
\begin{equation}
\mathcal{L}_{\mathrm{pre}}\!\Bigl(M,\,\bigl\{W^{(m)}\bigr\},\,\bigl\{f^{(m)}\bigr\}\Bigr)
\;=\;
\sum_{m=0}^{M-1}
\frac{W^{(m)}\,C}{f^{(m)}\,c},
\end{equation}
where $W^{(m)}$ denotes the mini-batch size in the $m$-th pre-training round, $f^{(m)}$ is the server’s CPU frequency, and $C$ is the number of FLOPs per sample. The constant $c$ represents additional overhead or cycle accuracy factors.

In Phase~II, each user $u\in\mathcal{U}$ trains locally on a mini-batch of size $b_{u}^{(n)}$ using the CPU frequency $f_{u}^{(n)}$. The local computation latency at round~$n$ is
\begin{equation}
\mathcal{L}_{u}^{(n),t}\bigl(b_{u}^{(n)}\bigr)
\;=\;
\frac{b_{u}^{(n)}\,C}{f_{u}^{(n)}\,c_{u}},
\end{equation}
where $c_{u}$ can be viewed as a user-specific overhead factor. In addition to local training, each user also incurs uplink and downlink latency for data transfer. Denoting respectively by $\mathcal{L}_{u}^{(n),d}$ and $\mathcal{L}_{u}^{(n),u}(p_{u}^{(n)})$ the one-way latencies associated with downlink and uplink transmission, the overall training latency for the two-stage system becomes
\begin{align}
&\tilde{\mathcal{L}}\Bigl(
  M,\,N,\,
  \{W^{(m)}\}, \{f^{(m)}\},
  \bigl\{b_{u}^{(n)}\bigr\}, \bigr\{f_{u}^{(n)}\bigr\}, \bigr\{p_{u}^{(n)}\bigr\},\bigr\{\boldsymbol{x}^{(n)}\bigr\},\nonumber\\
&\bigr\{\boldsymbol{q}^{(n)}\bigr\},\bigr\{\boldsymbol{w}^{(n)}\bigr\}
\Bigr)=\;
\mathcal{L}_{\mathrm{pre}}\!\Bigl(M,\{W^{(m)}\}, \{f^{(m)}\}\Bigr)
\;
\nonumber\\
&+\;
\sum_{n=0}^{N-1}
\max_{u\in \mathcal{U}}
\Bigl[
  \mathcal{L}_{u}^{(n),d}
  \;+\;
  \mathcal{L}_{u}^{(n),t}\bigl(b_{u}^{(n)}\bigr)
  \;+\;
  \mathcal{L}_{u}^{(n),u}\bigl(p_{u}^{(n)}\bigr)
\Bigr]
\nonumber\\
&=\;
\sum_{m=0}^{M-1} \frac{W^{(m)}\,C}{f^{(m)}\,c}
\;+\;
\sum_{n=0}^{N-1}
\max_{u\in \mathcal{U}}
\Bigl(
  \frac{\beta}{R_{u}^{(n),dl}}
  \;+\;
  \frac{b_{u}^{(n)}\,C}{f_{u}^{(n)}\,c_{u}}
  \;+\nonumber\\
&\;
  \frac{\beta}{R_{u}^{(n),ul}\bigl(p_{u}^{(n)}\bigr)}
\Bigr),
\end{align}
where $\beta$ is the payload size (in bits) for transmitting model updates, $R_{u}^{(n),dl}$ and $R_{u}^{(n),ul}\bigl(p_{u}^{(n)}\bigr)$ represent the respective downlink and uplink transmission rates (as in Section~\ref{subsec:comm}), and $p_{u}^{(n)}$ is the user’s transmit power at round~$n$.

\section{Joint Resource-Allocation Problem Formation}
\label{sec:problem}

We now present the unified optimization problem that underpins our proposed two-phase, wirelessly aware LLM training framework. The goal is to jointly allocate training resources, communication strategies, and physical-layer parameters in order to achieve efficient and accurate model convergence under realistic system constraints.

Let $M$ and $N$ denote the number of global rounds assigned to \emph{Phase~I} (centralized pre-training) and \emph{Phase~II} (federated fine-tuning), respectively. During Phase~I, the server selects a mini-batch size $W^{(m)}$ and CPU frequency $f^{(m)}$ in each round $m=0,1,\dots,M{-}1$. Similarly, in Phase~II, each user $u \in \mathcal{U}$ selects a local mini-batch size $b_{u}^{(n)}$ and a computation frequency $f_{u}^{(n)}$ for each round $n=0,1,\dots,N{-}1$.

Then, we employ a reconfigurable MA array at the server. In round $n$, the $i$-th antenna element is located at a spatial coordinate $\mathbf{x}_{n,i}\in\mathbb{R}^2$, and the receive beamforming vector is $\mathbf{q}^{(n)}\in\mathbb{C}^{N_T}$, constrained to a unit norm. The set of users participating in the round $n$ is indicated by $\mathcal{U}_n \subseteq \mathcal{U}$.

Thus, we denote by $\Psi(\cdot)$ the overall training objective, which aims to minimize the loss of fine-tuning of the LLM under time and energy constraints. Key constraints include:
\begin{itemize}
    \item \textbf{Time and Energy}: 
    $\mathcal{L}(M,N,\dots)$ and $Q(M,N,\dots)$ represent the total training time and energy usage in both phases, respectively.
    The detailed expression of the function can be found in\cite{Chen2021BeamformingOTA}.
    Each is limited by a practical budget, for example, $\mathcal{L}_{\max}$, $Q_{\max}$.
    \item \textbf{Antenna Spacing}: 
    A minimum gap $v > 0$ ensures $\|\mathbf{x}_{n,i} - \mathbf{x}_{n,j}\|\ge v$ for all $i,j$, preventing interference and coupling.
    \item \textbf{Beamforming Norm}: 
    $\|\mathbf{q}^{(n)}\|=1$ for each round $n$.
    \item \textbf{User Capacity}: 
    $|\mathcal{U}_n|\le U_{\max}$ imposes a limit on how many users can be scheduled in each round.
\end{itemize}

In summary, combining these components, we formulate the joint optimization problem $\mathbf{P1}$ (shown in problem~\ref{prob}). This problem integrates decisions between layers that span the geometry of the physical antenna, scheduling, and hyperparameters of machine learning. Because the selection of $\mathcal{U}_n$ directly influences the beamformer $\mathbf{q}^{(n)}$, and thus energy consumption, latency, and convergence speed, $\mathbf{P1}$ is naturally non-convex with integer and combinatorial aspects. Given its high dimensionality and the heterogeneous nature of its variables, we adopt an alternating optimization approach. This strategy partitions the problem into manageable subproblems dedicated to individual ``layers''.

\begin{align}\label{prob}
(\mathbf{P1}):
&\underset{
    \substack{
      M,\,N \in \mathbb{Z}_{+}\\
      \{W^{(m)},\,f^{(m)}\},\;\{b_{u}^{(n)}\},\{f_{u}^{(n)}\}\\
      \{\mathbf{x}^{(n)}\},\;\{\mathbf{q}^{(n)}\},\;\{\mathbf{w}^{(n)}\},\;\{\mathcal{U}_{n}\}
    }
}{\min}\;\Psi\Bigl(M,N,\;\{W^{(m)}\},\;\{b_{u}^{(n)}\}\Bigr)
\\
&\text{s.t.}\quad
\begin{aligned}[t]
&\mathcal{L}\Bigl(
  M,\,N,\,
  \{W^{(m)}\}, \{f^{(m)}\},
  \bigl\{b_{u}^{(n)}\bigr\}, \bigr\{f_{u}^{(n)}\bigr\}, \\&\qquad\bigr\{p_{u}^{(n)}\bigr\},
  \bigr\{\boldsymbol{x}^{(n)}\bigr\},\bigr\{\boldsymbol{q}^{(n)}\bigr\},\bigr\{\boldsymbol{w}^{(n)}\bigr\}
\Bigr)\le \mathcal{L}_{\max},
\end{aligned}\tag{P1a}
\\
&\qquad
\begin{aligned}[t]
&Q\Bigl(
  M,\,N,\,
  \bigl\{W^{(m)}\bigr\}, \bigl\{f^{(m)}\bigr\},
  \bigl\{b_{u}^{(n)}\bigr\}, \bigl\{f_{u}^{(n)}\bigr\}, \\&\qquad \bigl\{p_{u}^{(n)}\bigr\},
  \bigr\{\boldsymbol{x}^{(n)}\bigr\},\bigr\{\boldsymbol{q}^{(n)}\bigr\},\bigr\{\boldsymbol{w}^{(n)}\bigr\}
\Bigr)\le Q_{\max},
\end{aligned}\tag{P1b}
\\
&\qquad
\|x^{(n)}_{n_{1}} - x^{(n)}_{n_{2}}\| \ge v,
\quad \forall\,n_{1},n_{2}\in \mathcal{N}, \tag{P1c}
\\
&\qquad
\|\mathbf{q}^{(n)}\| = 1,\quad \forall\,n, \tag{P1d}
\\
&\qquad
\|\mathbf{w}^{(n)}\| = 1,\quad \forall\,n, \tag{P1e}
\\
&\qquad
\mathcal{U}_n \subseteq \mathcal{U},
\quad |\mathcal{U}_n| \le U_{\max},\quad \forall\,n. \tag{P1f}
\end{align}

\begin{algorithm}
\caption{Hybrid SCA--PDD Algorithm.}
\label{algo}
\begin{algorithmic}[1]
\State \textbf{Input:} Candidate training rounds $\mathcal{M}, \mathcal{N}$; constraints $\mathcal{L}_{\max}, Q_{\max}$; tolerances $\epsilon$, $\delta$; max iterations $T_{\max}$
\State \textbf{Initialize:} Best objective $\Psi^* \gets \infty$; optimal solution $\mathbf{z}^* \gets \emptyset$
\For{each $(M, N) \in \mathcal{M} \times \mathcal{N}$}, all optimization variables are partitioned into two sub-blocks.
    \State Initialize $\mathbf{z}_A^{(0)} = \{W^{(m)}, f^{(m)}, b_u^{(n)}, f_u^{(n)}\}$
    \State Initialize $\mathbf{z}_B^{(0)} = \{\mathbf{x}^{(n)}, \mathbf{q}^{(n)}, \mathbf{w}^{(n)}, \mathcal{U}_n\}$
    \For{$t = 1$ to $T_{\max}$}

        \State \textbf{Step 1: SCA Update of Continuous Variables $\mathbf{z}_A$}
        \ForAll{$m = 0,\dots,M{-}1$}
            \State Approximate $1/f^{(m)}$, $(f^{(m)})^3$ from latency/energy in Eq.~(25), Eq.~(34)
        \EndFor
        \ForAll{$n = 0,\dots,N{-}1$, $u \in \mathcal{U}$}
            \State Linearize $1/f_u^{(n)}$, $(f_u^{(n)})^2$ and $1/b_u^{(n)}$ from Eq.~(26), Eq.~(35)
        \EndFor
        \State Solve convex surrogate program for $\mathbf{z}_A^{(t)}$ under constraints:
        \begin{itemize}
            \item $\tilde{\mathcal{L}}(\cdot) \le \mathcal{L}_{\max}$ (Eq.~(25))
            \item $\tilde{Q}(\cdot) \le Q_{\max}$ (Eq.~(34))
            \item Bounds: $f^{(m)} \in [f_{\min}, f_{\max}]$, $b_u^{(n)} \in [b_{\min}, b_{\max}]$
            \item Optional: trust region $\|\mathbf{z}_A^{(t)} - \mathbf{z}_A^{(t{-}1)}\| \le \delta$
        \end{itemize}

        \State \textbf{Step 2: PDD Update of Geometry Variables $\mathbf{z}_B$}
        \State Fix $\mathbf{z}_A^{(t)}$, and repeat the following until convergence:
        \State \quad (a) \textbf{Auxiliary Variables:} Solve convex problem for $t_1, t_2, g_u^{ul}, g_u^{dl}$ (Eq.~(29)--(31))
        \State \quad (b) \textbf{Unit-Modulus Projection:}
        \State \qquad Solve for $\boldsymbol{\theta}_u^{ul}, \boldsymbol{\theta}_u^{dl}$ s.t.\ $|\theta_i| = 1$ (Eq.~(32)--(33)) using \cite{cai2019robust}
        \State \quad (c) \textbf{Beamforming:} Optimize $\mathbf{q}^{(n)}, \mathbf{w}^{(n)}$ (Eq.~(31)) via oblique manifold solver~\cite{8594607}
        \State \quad (d) \textbf{MA Positioning:}
        \State \qquad Solve:
        \[
        \min_{\mathbf{x}^{(n)}} \sum_{u} \|\angle \boldsymbol{\theta}_u - \angle \boldsymbol{a}_u(\mathbf{x}^{(n)})\|^2
        \]
        \qquad subject to spacing: $\|\mathbf{x}_{i} - \mathbf{x}_{j}\| \ge v$ using SCA (Eq.~(37)--(39))
        \State \quad (e) \textbf{Penalty Update:} If spacing or rate constraints violated, update dual variables

        \If{$\|\mathbf{z}^{(t)} - \mathbf{z}^{(t{-}1)}\| < \epsilon$}
            \State \textbf{break}
        \EndIf
    \EndFor
    \If{$\Psi(M,N) < \Psi^*$}
        \State Update $\Psi^* \gets \Psi(M,N)$ and store solution $\mathbf{z}^*$
    \EndIf
\EndFor
\State \textbf{Return:} Optimal $(M^*, N^*, \mathbf{z}_A^*, \mathbf{z}_B^*)$
\end{algorithmic}
\end{algorithm}

\section{Hybrid SCA--PDD Algorithm}
\label{sec:Algorithm}

To address the above problem, we propose a hybrid SCA-PDD algorithm as shown in Algorithm~\ref{algo}.  At the outer level, we enumerate every feasible integer pair \((M,N)\) that satisfies the overall latency-energy envelope. For each such pair, an inner loop alternates between two blocks: (i) an SCA step that linearizes and optimizes smooth continuous variables and (ii) a PDD step that resolves the discrete user selection decisions and the geometry constraints of the MA array. The two blocks iterate until the joint objective converges, after which the best solution among all the enumerated pairs \((M,N)\) is retained as the final allocation of resources.

\subsection{Enumerating \boldmath$(M,N)$}

Because \(M\) and \(N\) are integers and typically have an upper bound in practice, we can scan all pairs \(\bigl(M,N\bigr)\) in a discrete set. For each pair, we fix \(M\) and \(N\) in problem \(\mathbf{P1}\) and solve the resulting continuous/discrete optimization in the remaining variables. Among all enumerated pairs, we select the one that yields the best objective \(\Psi(\cdot)\).

\subsection{Partitioning Variables for SCA vs.\ PDD}

Let
$
  \mathbf{z} \;=\; 
  \Bigl\{
    W^{(m)}, f^{(m)},\, b_{u}^{(n)}, f_{u}^{(n)},\,
    \mathbf{x}^{(n)},\, \mathbf{q}^{(n)},\, \mathbf{w}^{(n)},\,
    \mathcal{U}_n
  \Bigr\}.
$
Then, we split \(\mathbf{z}\) into two blocks:
$
  \mathbf{z}_A \quad (\text{SCA block})\quad\text{and}\quad
  \mathbf{z}_B \quad (\text{PDD block}).
$
In particular:
\begin{itemize}
  \item \(\mathbf{z}_A\) might include continuous variables that enter the objective or constraints in a smooth but non-convex manner (e.g. functions \(1/W^{(m)}\) and \(\kappa_0 (f^{(m)})^3\)).
  \item \(\mathbf{z}_B\) contains discrete/user-selection decisions \(\{\mathcal{U}_n\}\), MA geometry \(\{\mathbf{x}_{n,i}\}\) subject to minimum spacing constraints, and beamforming vectors \(\{\mathbf{q}^{(n)}\}\) with \(\|\mathbf{q}^{(n)}\|=1\). These typically require specialized handling.
\end{itemize}

\paragraph{Iterative Procedure}
\begin{enumerate}
    \item \textbf{Block A: SCA Step.}
      Fix the current values of \(\mathbf{z}_B\) (user selections, antenna coordinates, beamforming).  Then linearize each non-convex but differentiable term in \(\mathbf{z}_A\) around the old iterate, leading to a convex approximation.  Solve this convex subproblem to update \(\mathbf{z}_A\).
    \item \textbf{Block B: PDD Step.}
      Fix the updated \(\mathbf{z}_A\). Handle discrete variables (\(\mathcal{U}_n\)) and geometry constraints (e.g.\ \(\|\mathbf{x}_{n,i}-\mathbf{x}_{n,j}\|\!\ge v\)) using a partial Lagrangian or penalty approach. Update the dual multipliers for time/energy constraints and other global coupling constraints.
\end{enumerate}
This two-block iteration repeats until convergence (e.g., changes in \(\mathbf{z}\) fall below a tolerance).  Because \((M,N)\) are fixed in each run, these steps address only the continuous and discrete variables in \(\mathbf{z}_A,\mathbf{z}_B\).

\subsubsection*{Block A: SCA Details}
Any non-convex function in \(\mathbf{z}_A\) that is differentiable can be replaced by a first-order Taylor approximation around the previous iteration.  
For example:
\begin{equation}
  \frac{1}{W^{(m)}} 
  \;\approx\;
  \frac{1}{W^{(m)}_{\text{old}}}
  \;-\;
  \frac{W^{(m)} - W^{(m)}_{\text{old}}}{\bigl(W^{(m)}_{\text{old}}\bigr)^2}.
\end{equation}
Similarly, \(\kappa_0(f^{(m)})^3\) is linearized around \(f^{(m)}_{\text{old}}\).  Substituting these linear approximations into \(\mathbf{P1}\) yields a convex subproblem solvable by standard methods.

\subsubsection{Block B: PDD Details}
In this block, we optimize the transmit beamforming 
$\boldsymbol{w}^{(n)}$, the receive beamforming $\boldsymbol{q}^{(n)}$, and the MA position of the base station $\boldsymbol{x}^{(n)}$. First, we introduce the auxiliary variables $t_{1,u}$ and $t_{2,u}$. Based on the relaxation method, we have the following formulations
\begin{align}
&\frac{\beta}{B_{dl}\log_{2}(1+\frac{\boldsymbol{h}^{H}_{dl,u}(\boldsymbol{x}^{(n)})\boldsymbol{w}^{(n)}\tilde{P}}{B_{dl}\tilde{N}_{dl,u}^{(n)}})}\leq t_{1},\\
&\frac{\beta}{B_{ul}^{u}\log_{2}(1+\frac{(\boldsymbol{q}^{(n)})^{H}\boldsymbol{h}_{ul}^{u}(\boldsymbol{x}^{(n)})p_{u}^{(n)}}{B_{ul}^{u}N_{ul}^{n}})}\leq t_{2}.
\end{align}
The two terms 
$(\boldsymbol{q}^{(n)})^{H}\boldsymbol{h}_{ul}^{u}(\boldsymbol{x}^{(n)})$ and $\boldsymbol{h}_{dl,u}^{u}(\boldsymbol{x}^{(n)})\boldsymbol{w}^{(n)}$
exhibit a complex interdependency. Moreover, the channel model, as presented in equations (11) and (14), can be reformulated as
\begin{align}
&(\boldsymbol{q}^{(n)})^{H}\boldsymbol{h}_{ul}^{u}(\boldsymbol{x}^{(n)})=\alpha_{u}^{(n),l}(\boldsymbol{q}^{(n)})^{H}\boldsymbol{a}_{u}^{ul}(\boldsymbol{x}^{(n)}),\nonumber\\
&\boldsymbol{h}_{dl,u}^{H}(\boldsymbol{x}^{(n)})\boldsymbol{w}^{(n)}=(\beta_{u}^{(n)})^{H}(\boldsymbol{a}_{u}^{dl}(\boldsymbol{x}^{(n)}))^{H}\boldsymbol{w}^{(n)}.
\end{align}
Since $\alpha_{u}^{(n),l}$ and $\beta_{u}^{(n)}$
are known, and the receive beamforming vector $\boldsymbol{q}^{(n)}$ is coupled with the receive array response vector $\boldsymbol{a}_{u}^{ul}(\boldsymbol{x}^{(n)})$, while the transmit beamforming vector $\boldsymbol{w}^{(n)}$
is coupled with $\boldsymbol{a}_{u}^{dl}(\boldsymbol{x}^{(n)})$, we introduce auxiliary variables $g_{u}^{(n),ul}$ and $g_{u}^{(n),dl}$ to decouple the system, and the channel can be reformulated using substituted variables
\begin{align}
&g_{u}^{(n),ul}=(\boldsymbol{q}^{(n)})^{H}\boldsymbol{a}_{u}^{ul}(\boldsymbol{x}^{(n)}),\nonumber\\
&g_{u}^{(n),dl}=(\boldsymbol{a}_{u}^{dl}(\boldsymbol{x}^{(n)}))^{H}\boldsymbol{w}^{(n)}.
\end{align}
Due to the coupling of antenna position variables in the array response vectors 
$\boldsymbol{a}_{u}^{ul}(\boldsymbol{x}^{(n)})$ and $\boldsymbol{a}_{u}^{dl}(\boldsymbol{x}^{(n)})$, and the fact that both of these response vectors satisfy a constant modulus constraint, we introduce auxiliary variables $\boldsymbol{\theta}_{u}^{ul}=\boldsymbol{a}_{u}^{ul}(\boldsymbol{x}^{(n)})$ and $\boldsymbol{\theta}_{u}^{dl}=\boldsymbol{a}_{u}^{dl}(\boldsymbol{x}^{(n)})$, which also satisfy the constant modulus constraint. Consequently, we impose the following new constraints
\begin{align}
&\boldsymbol{\theta}_{u}^{ul}=\boldsymbol{a}_{u}^{ul}(\boldsymbol{x}^{(n)}), |\boldsymbol{\theta}_{u}^{ul}(i)|=1, \nonumber\\
&\boldsymbol{\theta}_{u}^{dl}=\boldsymbol{a}_{u}^{dl}(\boldsymbol{x}^{(n)}), |\boldsymbol{\theta}_{u}^{dl}(i)|=1.
\end{align}
By introducing the auxiliary variables and constraints, Problem A can be rewritten as follows.
\begin{subequations}
\begin{align}
\min_{\boldsymbol{\theta}_{u}^{ul},\boldsymbol{\theta}_{u}^{dl},t_{1},t_{2},g_{u}^{(n),ul},\atop g_{u}^{(n),dl},\boldsymbol{x}^{(n)},\boldsymbol{q}^{(n)},\boldsymbol{w}^{(n)}}&~\Psi(M,N,\{W^{(m)}\},\{b_{u}^{(n)}\})\label{pro16a}\\
\mbox{s.t.}~
&\sum_{m=0}^{M-1}\frac{W^{(m)}C}{f^{(m)}c}+\sum_{n=0}^{N-1}(t_{1}+t_{2}+\nonumber\\
&\max_{k\in\mathcal{K}}(\frac{b_{u}^{(n)}C}{\hat{f}_{u}^{(n)}c_{u}}))\leq\tilde{\mathcal{L}}_{0},&\\
&\sum_{m=0}^{M-1}\eta\frac{W^{(m)C}}{c}\phi[f^{(m)}]^{2}+\sum_{n=0}^{N-1}\tilde{P}t_{1}+\nonumber\\
&\sum_{n=0}^{N-1}\sum_{u=1}^{U}(Q_{u}^{(n),t}(b_{u}^{(n)},f_{u}^{(n)})+p_{u}^{(n)}t_{2})\leq\tilde{E}_{0},&\\
&\boldsymbol{\theta}_{u}^{ul}=\boldsymbol{a}_{u}^{ul}(\boldsymbol{x}^{(n)}), |\boldsymbol{\theta}_{u}^{ul}(i)|=1,&\\
&\boldsymbol{\theta}_{u}^{dl}=\boldsymbol{a}_{u}^{dl}(\boldsymbol{x}^{(n)}), |\boldsymbol{\theta}_{u}^{dl}(i)|=1,&\\
&\frac{\beta}{B_{dl}\log_{2}(1+\frac{g_{u}^{(n),dl}(\beta_{u}^{(n)})^{H}\tilde{P}}{B_{dl}\tilde{N}_{dl,u}^{(n)}})}\leq t_{1},&\\
&\frac{\beta}{B_{ul}^{u}\log_{2}(1+\frac{g_{u}^{(n),ul}\alpha_{u}^{(n),l}p_{u}^{(n)}}{B_{ul}^{u}N_{ul}^{n}})}\leq t_{2},&\\
&g_{u}^{(n),ul}=(\boldsymbol{q}^{(n)})^{H}\boldsymbol{\theta}_{u}^{ul},&\\
&g_{u}^{(n),dl}=(\boldsymbol{\theta}_{u}^{dl})^{H}\boldsymbol{w}^{(n)}.
\end{align}\label{pro16}
\end{subequations}

Therefore, in this section, the optimization variables $\{\boldsymbol{\theta}_{u}^{ul}$,$\boldsymbol{\theta}_{u}^{dl}$,$t_{1}$,$t_{2}$,$g_{u}^{(n),ul}$,$ g_{u}^{(n),dl}$,$\boldsymbol{x}^{(n)}$,$\boldsymbol{q}^{(n)}$,$\boldsymbol{w}^{(n)}\}$ can be divided into four sub-problems. The first sub-problem involves the variables $\{t_{1},t_{2},g_{u}^{(n),ul}, g_{u}^{(n),dl}\}$, the second sub-problem involves the variables $\{\boldsymbol{\theta}_{u}^{ul},\boldsymbol{\theta}_{u}^{dl}\}$, the third sub-problem involves the variables $\{\boldsymbol{q}^{(n)},\boldsymbol{w}^{(n)}\}$, and the fourth sub-problem involves the variable $\{\boldsymbol{x}^{(n)}\}$. Next, we provide details on how to solve these four subproblems.

\textbf{The first sub-problem over} $\{t_{1},t_{2},g_{u}^{(n),ul}, g_{u}^{(n),dl}\}$: Problem B is convex with respect to the four auxiliary variables, and thus we can use CVX to solve the first subproblem.

\textbf{The second sub-problem over} $\{\boldsymbol{\theta}_{u}^{ul},\boldsymbol{\theta}_{u}^{dl}\}$: The optimization problem involving the constraints on variables
$\boldsymbol{\theta}_{u}^{ul}$ and $\boldsymbol{\theta}_{u}^{dl}$ can be reformulated as follows:
\begin{subequations}
\begin{align}
\min_{\boldsymbol{q}_{d,k}^{(n)},\boldsymbol{q}_{k}^{u}}&~\sum_{u=1}^{U}|g_{u}^{(n),ul}-(\boldsymbol{q}^{(n)})^{H}\boldsymbol{\theta}_{u}^{ul}|^{2}+|g_{u}^{(n),dl}-(\boldsymbol{\theta}_{u}^{dl})^{H}\boldsymbol{w}^{(n)}|^{2}\nonumber\\
&+\|\boldsymbol{\theta}_{u}^{ul}-\boldsymbol{a}_{u}^{ul}(\boldsymbol{x}^{(n)})\|^{2}+
\|\boldsymbol{\theta}_{u}^{dl}-\boldsymbol{a}_{u}^{dl}(\boldsymbol{x}^{(n)})\|^{2},&\label{pro16a}\\
\mbox{s.t.}~
&|\boldsymbol{\theta}_{u}^{ul}(i)|=1,&\\
&|\boldsymbol{\theta}_{u}^{dl}(i)|=1.
\end{align}\label{pro16}
\end{subequations}
Due to non-convex constraints $\|\boldsymbol{q}_{d,k}^{(n)}\|=1$ and $\|\boldsymbol{q}_{k}^{u,n}\|=1$, solving this problem is challenging. However, the algorithm proposed in \cite{cai2019robust}, one iteration block coordinate descent type algorithm, can address the above non-convex problem. The details of this algorithm are omitted here.

\textbf{The third sub-problem over} $\{\boldsymbol{w}^{(n)},\boldsymbol{q}^{(n)}\}$: The optimization problem involving the constraints on variables 
$\boldsymbol{w}^{(n)}$ and 
$\boldsymbol{q}^{(n)}$ can be reformulated as follows: 
\begin{subequations}
\begin{align}
\begin{split}
\min_{\boldsymbol{q}_{d,k}^{(n)},\boldsymbol{q}_{k}^{u}}~ 
&\sum_{u=1}^{U} \Bigl(
|g_{u}^{(n),ul} - (\boldsymbol{q}^{(n)})^{H} \boldsymbol{\theta}_{u}^{ul}|^{2} \\
&\quad + |g_{u}^{(n),dl} - (\boldsymbol{\theta}_{u}^{dl})^{H} \boldsymbol{w}^{(n)}|^{2} \Bigr)
\end{split} \label{pro16a}\\
\mbox{s.t.}~\quad
&\|\boldsymbol{w}^{(n)}\|^{2} = 1, \\
&\|\boldsymbol{q}^{(n)}\|^{2} = 1.
\end{align}\label{pro16}
\end{subequations}
Because non-convex constraints $\|\boldsymbol{w}^{(n)}\|^{2}=1$ and $\|\boldsymbol{q}^{(n)}\|^{2}=1$, solving this problem is challenging. However, the algorithm proposed in \cite{8594607}, an oblique manifold algorithm, can address the above non-convex problem. The details of the oblique manifold algorithm can be found in \cite{8594607}.

\textbf{The fourth sub-problem over} $\{\boldsymbol{x}^{(n)}\}$: From the previous steps, we can determine the values of the auxiliary variables $\boldsymbol{\theta}_{u}^{ul}$ and $\boldsymbol{\theta}_{u}^{dl}$. To satisfy the equality constraints 
$\boldsymbol{\theta}_{u}^{ul}=\boldsymbol{a}_{u}^{ul}(\boldsymbol{x}^{(n)})$ and $\boldsymbol{\theta}_{u}^{dl}=\boldsymbol{a}_{u}^{dl}(\boldsymbol{x}^{(n)})$, we have
\begin{subequations}
\begin{align}
\min_{\boldsymbol{x}^{(n)}}~&\sum_{u=1}^{U}\|\angle\boldsymbol{\theta}_{u}^{ul}-\angle\boldsymbol{a}_{u}^{ul}(\boldsymbol{x}^{(n)})\|^{2}+
\|\angle\boldsymbol{\theta}_{u}^{dl}-\angle\boldsymbol{a}_{u}^{dl}(\boldsymbol{x}^{(n)})\|^{2}&\\
&\|x_{n_{1}}^{(n)}-x_{n_{2}}^{(n)}\|\geq v.
\end{align}    
\end{subequations}
To solve for the MA position, we further reformulate the problem using matrix norm operations as follows
\begin{subequations}
\begin{align}
\begin{split}
\min_{\boldsymbol{x}^{(n)}}~&\sum_{u=1}^{U} \sum_{n_{1}=1}^{N_{t}} \Bigl(
|\angle\boldsymbol{\theta}_{u}^{ul}(n_{1}) - x_{n_{1}}^{(n)}\cos(\phi_{u})|^{2} \\
&\quad + |\angle\boldsymbol{\theta}_{u}^{ul}(n_{1}) - x_{n_{1}}^{(n)}\sin(\theta_{u})|^{2} \Bigr)
\end{split} \\
&|x_{n_{1}}^{(n)} - x_{n_{2}}^{(n)}| \geq v.
\end{align}
\end{subequations}
We assume that $x_{n_{2}}^{(n)}$ is the initial MA position. Next, we apply the SCA algorithm to solve for 
$x_{n_{1}}^{(n)}$. Using the SCA algorithm, the constraint 
$|x_{n_{1}}^{(n)}-x_{n_{2}}^{(n)}|\geq v$ can be transformed into
\begin{align}
\sqrt{(x_{n_1}^{(n)} - x_{n_{2}}^{(n)})^2} + \frac{x_{n_1}^{(n)} - x_{n_{2}}^{(n)}}{|x_{n_1}^{(n)} - x_{n_{2}}^{(n)}|}(x_{n_1}^{(n)} - x_{n_{2}}^{(n)})\geq v.
\end{align}
The problem (37) is then reformulated as
\begin{subequations}
\begin{align}
\begin{split}
\min_{\boldsymbol{x}^{(n)}}~\sum_{u=1}^{U}\sum_{n_{1}=1}^{N_{t}} \Bigl(
&|\angle\boldsymbol{\theta}_{u}^{ul}(n_{1}) - x_{n_{1}}^{(n)}\cos(\phi_{u})|^{2} \\
&\quad + |\angle\boldsymbol{\theta}_{u}^{ul}(n_{1}) - x_{n_{1}}^{(n)}\sin(\theta_{u})|^{2} \Bigr)
\end{split} \\
\begin{split}
\sqrt{(x_{n_1}^{(n)} - x_{n_{2}}^{(n)})^2}
&+ \frac{x_{n_1}^{(n)} - x_{n_{2}}^{(n)}}{|x_{n_1}^{(n)} - x_{n_{2}}^{(n)}|}(x_{n_1}^{(n)} - x_{n_{2}}^{(n)}) \\
&\geq v.
\end{split}
\end{align}
\end{subequations}
Finally, we observe that the problem with respect to the MA position 
$x_{n,1}^{(n)}$ is convex. Therefore, we apply CVX to solve the aforementioned problem. The complexity and convergence analysis are in Appendix~\ref{sec:Complexity} and Appendix~\ref{sec:Convergence}.

\section{Numerical Results}
\label{sec:num-results}

This section presents a detailed numerical study that is tailored to instruction-tuning benchmarks. Unless otherwise noted, we average results over ten independent runs of the wireless channel, with shaded areas in the figures indicating standard deviations.

\subsection{Experiment Setup}
\noindent\textbf{Experiment Settings.} All experiments are conducted on a server with Ubuntu~22.04, Python~3.12, CUDA~12.8 and PyTorch~2.8.0. The hardware consists of one NVIDIA L20 GPU (48~GB VRAM), an Intel Xeon Platinum~8457C CPUs (20~vCPUs), 100~GB system memory and a 500~GB disk. In the federated fine-tuning simulation, we emulate 10 clients \emph{using the Flower framework}~\cite{beutel2022flowerfriendlyfederatedlearning}. On the wireless side, we use a Rician channel with carrier frequency \(f_c=28\,\mathrm{GHz}\) and bandwidth \(B=20\,\mathrm{MHz}\) (K-factor \(=8\,\mathrm{dB}\), number of paths \(=3\), \(P_{\mathrm{LOS}}=0.8\), blockage probability \(=0.03\), shadowing standard deviation \(=4\,\mathrm{dB}\), user speed \(=0.2\,\mathrm{m/s}\)). The transmit-power limit is \(P_{\max}=0.2\,\mathrm{W}\), and the noise power spectral density is \(N_0=-174\,\mathrm{dBm/Hz}\). Each client link is equipped with a 16-element reconfigurable antenna. We set \texttt{fraction-fit}=1.0.

\noindent\textbf{Model Configuration.} We adopt the 3B-parameter OpenLLaMA~v2 model~\cite{openlm2023openllama}, applying 4-bit quantization via \texttt{bitsandbytes} (v0.45.4)~\cite{10.5555/3666122.3666563} and LoRA (\texttt{peft})~\cite{hu2021loralowrankadaptationlarge} for parameter-efficient fine-tuning (LoRA rank\,=\,32, $\alpha=64$, dropout\,=\,0.075).
To reduce GPU memory, we enable gradient checkpointing.
We use HuggingFace's \texttt{SFTTrainer}; the maximum sequence length is 512 and the learning rate is $5\times10^{-5}$.

\subsection{Baselines and Metrics}
\noindent\textbf{Baselines.} We compare five schedulers :

\textit{Digital FedAvg (OFDMA).} Orthogonal uplink with (assumed) error-free digital aggregation; all selected clients upload over orthogonal resource blocks (RBs).

\textit{TopK-SNR.} Each round selects the $k$ clients with the highest instantaneous post-combining SNR at the BS; uplink aggregation is digital.

\textit{Gibbs.} Probabilistic client selection: sample $k$ clients with probability proportional to normalized link quality (no power control); aggregation is digital.

\textit{OTA No-PC.} Analog over-the-air (AirComp) aggregation with a \emph{fixed} BS array and \emph{no} per-user power control; $k$ clients are superposed each round. \textit{When $k{=}U$ (all clients), this recovers the commonly used “Select-All” variant.}

\textit{MA (Greedy).} Same OTA setting as above, but the BS movable/reconfigurable array geometry and receive beamformer are greedily updated each round; the client set follows TopK-SNR.

\noindent\textbf{Metrics.}  We report four metrics to capture \emph{fairness}, \emph{inequality}, \emph{model quality}, and \emph{radio link} as follows.

\textit{Fairness.}
Fairness is measured by the Jain index over per-client participation shares \(x_u\) :
\begin{equation}
\label{eq:jain}
J=\frac{\big(\sum_{u=1}^{U} x_u\big)^2}{U\sum_{u=1}^{U} x_u^2}, \qquad J\in(0,1].
\end{equation}
\(J=1\) indicates perfectly equal participation; lower values imply greater inequality.

\textit{Inequality.}
Inequality is measured by the Gini coefficient over per-client participation shares \(x_u\):
\begin{equation}
\label{eq:gini}
G=\frac{\sum_{u=1}^{U}\sum_{v=1}^{U}\lvert x_u-x_v\rvert}{2U\sum_{u=1}^{U}x_u},\qquad G\in[0,1).
\end{equation}
Lower \(G\) indicates more equal participation and complements the Jain index.

\textit{Model quality.}
The quality of the model at the R30 horizon is reported as perplexity
\begin{equation}
\label{eq:ppl-r30}
\mathrm{PPL}(\mathrm{R30})=\exp\!\big(\ell(\boldsymbol{\theta}_{\mathrm{R30}})\big),
\end{equation}
where \(\ell\) is the average validation NLL per token (use \(2^{\ell}\) if NLL is in bits). Lower PPL implies a better predictive fit / faster convergence by R30.

\textit{Link quality and efficiency.}
We report the average post-combining SNR at the BS in dB:
\begin{equation}
\label{eq:avg-snr}
\mathrm{Avg\,SNR(dB)}=\frac{1}{N_{\mathrm{R30}}}\sum_{n=1}^{N_{\mathrm{R30}}}10\log_{10}\!\big(\mathrm{SNR}^{(n)}\big).
\end{equation}
Higher is generally better; under analog OTA aggregation, it must be read alongside the analogue‐sum mismatch.

\begin{figure*}
    \centering
    \includegraphics[width=\linewidth]{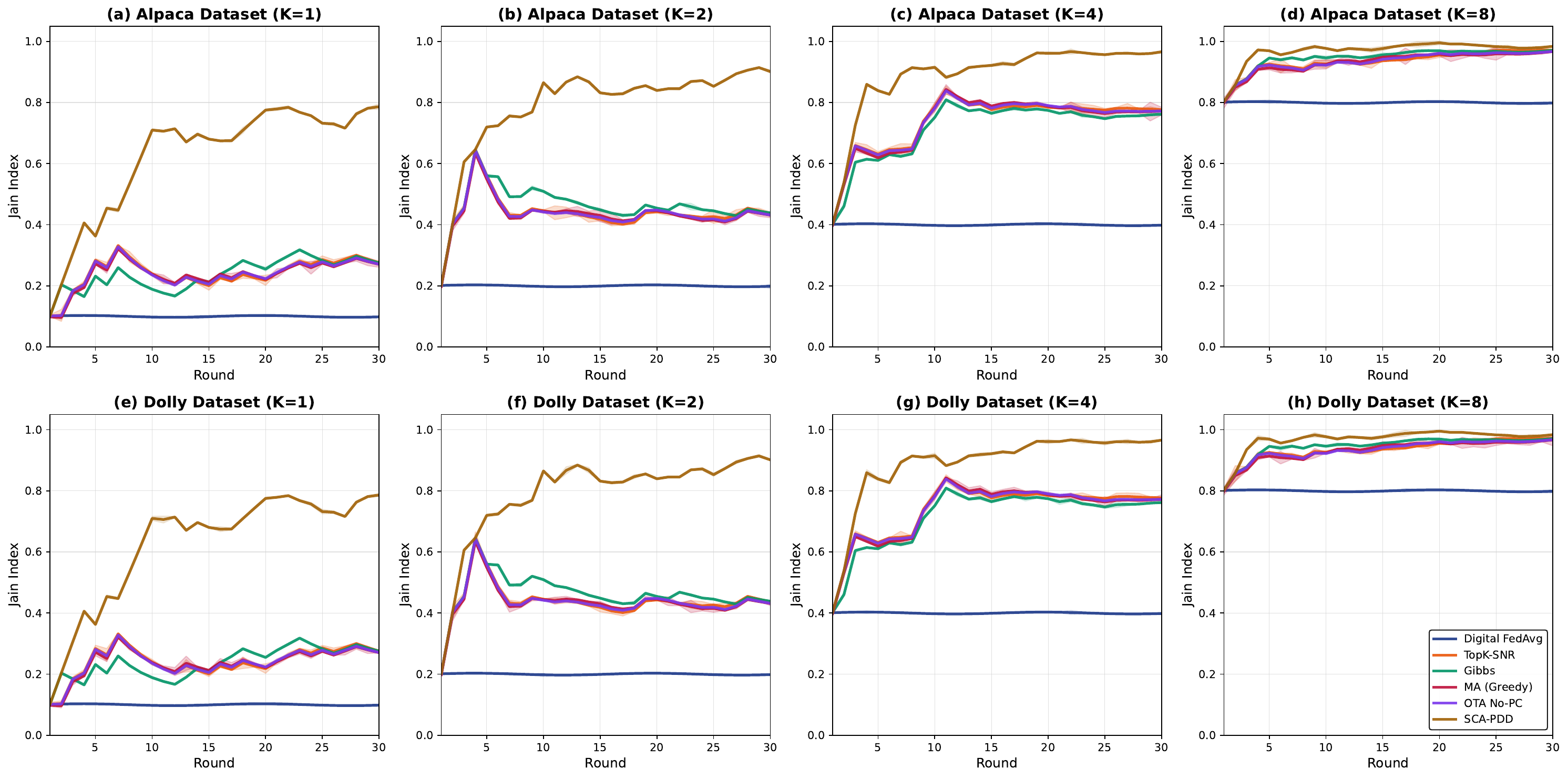}
    \caption{Jain Index comparison across different learning methods on Alpaca and Dolly datasets. (a)-(d) show results for Alpaca dataset with K=1,2,4,8 clients per round, while (e)-(h) show corresponding results for Dolly dataset. Shaded regions represent method-specific variations to distinguish overlapping performance curves.}
    \label{fig:jain}
\end{figure*}

\subsection{Alpaca Experiments}
\label{sec:alpaca}

The \textsc{Alpaca} corpus (52\,k instruction–response pairs) \cite{taori2023alpaca} is evenly sharded into ten non‑overlapping subsets of ${\approx}5{,}200$ samples.

\subsubsection{Fairness}
SCA–PDD achieves the highest Jain index at all concurrency levels $k$ as shown in Fig.~\ref{fig:jain} (a)-(d). At $k{=}1$ the index is 0.789 versus 0.285 for the best baseline (Gibbs), an absolute gain of 0.504; at $k{=}2$ it is 0.905 versus 0.449, a gain of 0.456. With larger concurrency, fairness rises for all methods because more clients are active per round, yet SCA–PDD remains ahead: 0.969 versus 0.774 at $k{=}4$ (gain 0.195) and 0.985 versus 0.972 at $k{=}8$ (gain 0.013). These outcomes reflect that SNR‑centric schedulers tend to concentrate airtime on a few strong links, whereas SCA–PDD couples client selection with beamforming and movable antenna geometry to keep the analog superposition well aligned while rotating participation, thereby broadening inclusion without compromising OTA stability, especially at small $k$ where fairness is most challenging.

\begin{figure*}
    \centering
    \includegraphics[width=\linewidth]{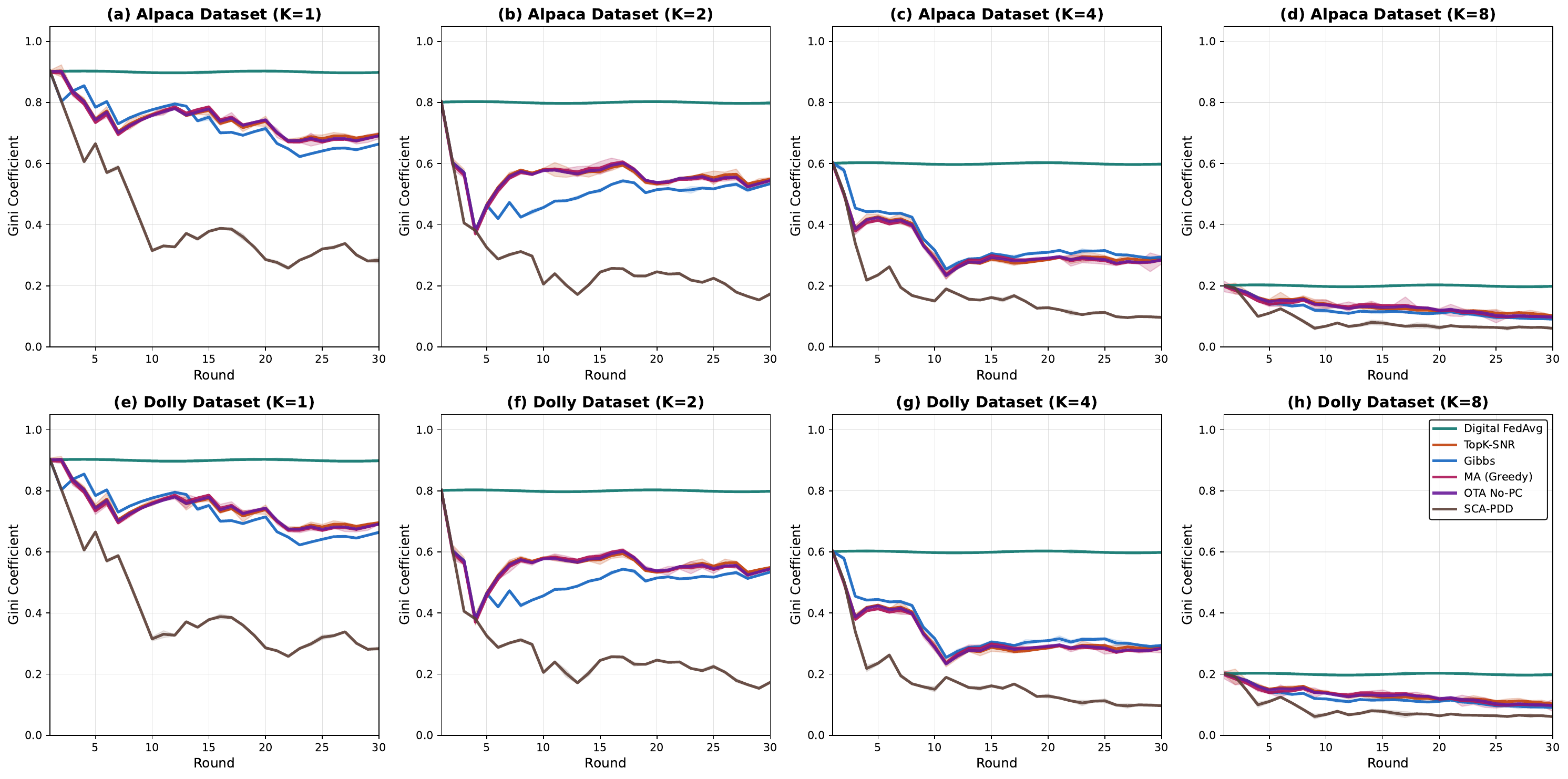}
    \caption{Gini coefficient evolution across different learning methods over 30 training rounds. Subplots (a)-(d) show results on Alpaca dataset and (e)-(h) on Dolly dataset with K=1,2,4,8 clients selected per round. Shaded regions represent method-specific variations for visual distinction.}
    \label{fig:gini}
\end{figure*} 

\subsubsection{Inequality}
The Gini coefficient corroborates the fairness gains and highlights how participation concentrates in the tail as illustrated in Fig.~\ref{fig:gini} (a)-(d). With $k{=}1$, SCA–PDD attains 0.287 compared to 0.640 for the best baseline (Gibbs), an absolute decrease of 0.353 (55.2\%). At $k{=}2$, the coefficient falls to 0.177 versus 0.531 (decrease 0.354; 66.7\%). The advantage persists at $k{=}4$ with 0.100 versus 0.287 (decrease 0.187; 65.2\%) and remains visible at $k{=}8$ with 0.069 versus 0.092 (decrease 0.023; 25.0\%). The largest reductions occur at small $k$, where SNR‑driven policies most strongly concentrate airtime. By rotating clients while adapting beamforming and movable antenna geometry to maintain a well‑aligned analog superposition, SCA–PDD shrinks the participation tail, weak links appear more often without destabilizing aggregation, consistent with the scheduling constraint in (P1f).

\begin{figure}[t]
  \centering
\includegraphics[width=\columnwidth]{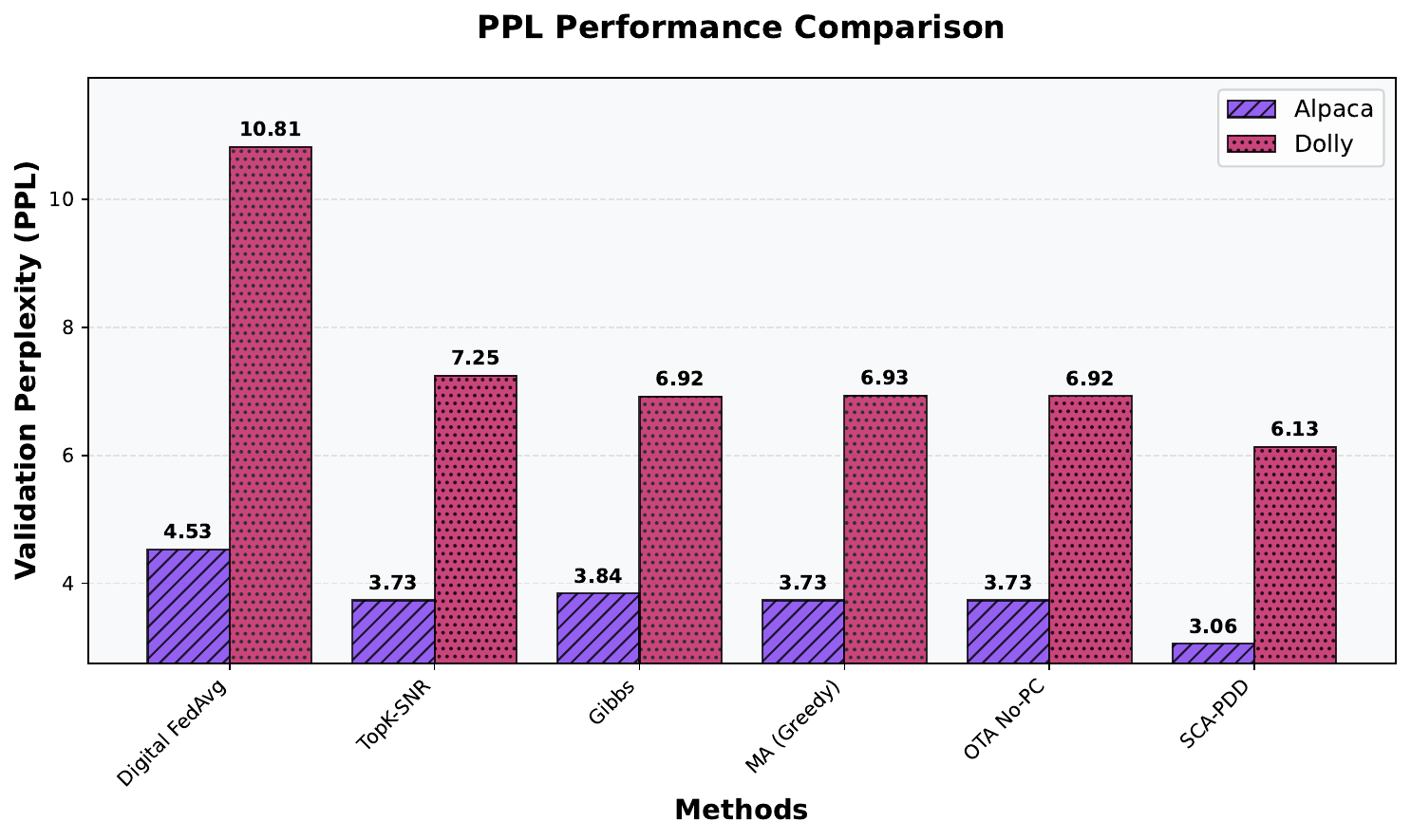}%
  \caption{PPL performance comparison across different learning methods on Alpaca and Dolly datasets at round 30.}
  \label{fig:ppl}
\end{figure}

\subsubsection{Model quality}
In a fixed round budget, the clearest gains appear in small–moderate concurrency (Fig.~\ref{fig:ppl}).  For $k{=}1$, the perplexity is 2.94 compared to 3.56 for the best baseline (Digital), an absolute reduction of 0.62 (17.4\%). 
For $k{=}2$, the result is 3.03 versus 3.53, a reduction of 0.50 (14.2\%). 
These improvements coincide with the strongest fairness gains and are consistent with the analysis that the analog‑sum mismatch term, rather than SNR alone, dominates learning quality when a few clients superpose.

When concurrency grows, alignment becomes intrinsically harder. At $k{=}4$, SCA–PDD attains 3.15 versus 3.12 (Gibbs), a difference of 0.03 (0.96\%), while requiring substantially less energy up to R30 (252\,J versus 364\,J; 30.8\% reduction), indicating a stronger quality–energy trade‑off. 
At $k{=}8$, perplexity is 3.97 versus 3.17 for the best baseline, reflecting the difficulty of perfectly aligning many simultaneous transmissions. 
Overall, the results point to small–moderate $k$ as a practical operating region where geometry‑aware scheduling most effectively translates into better fixed‑budget model quality.

\subsubsection{Link quality and efficiency}

\begin{figure}[t]
  \centering
  \includegraphics[width=0.7\columnwidth]{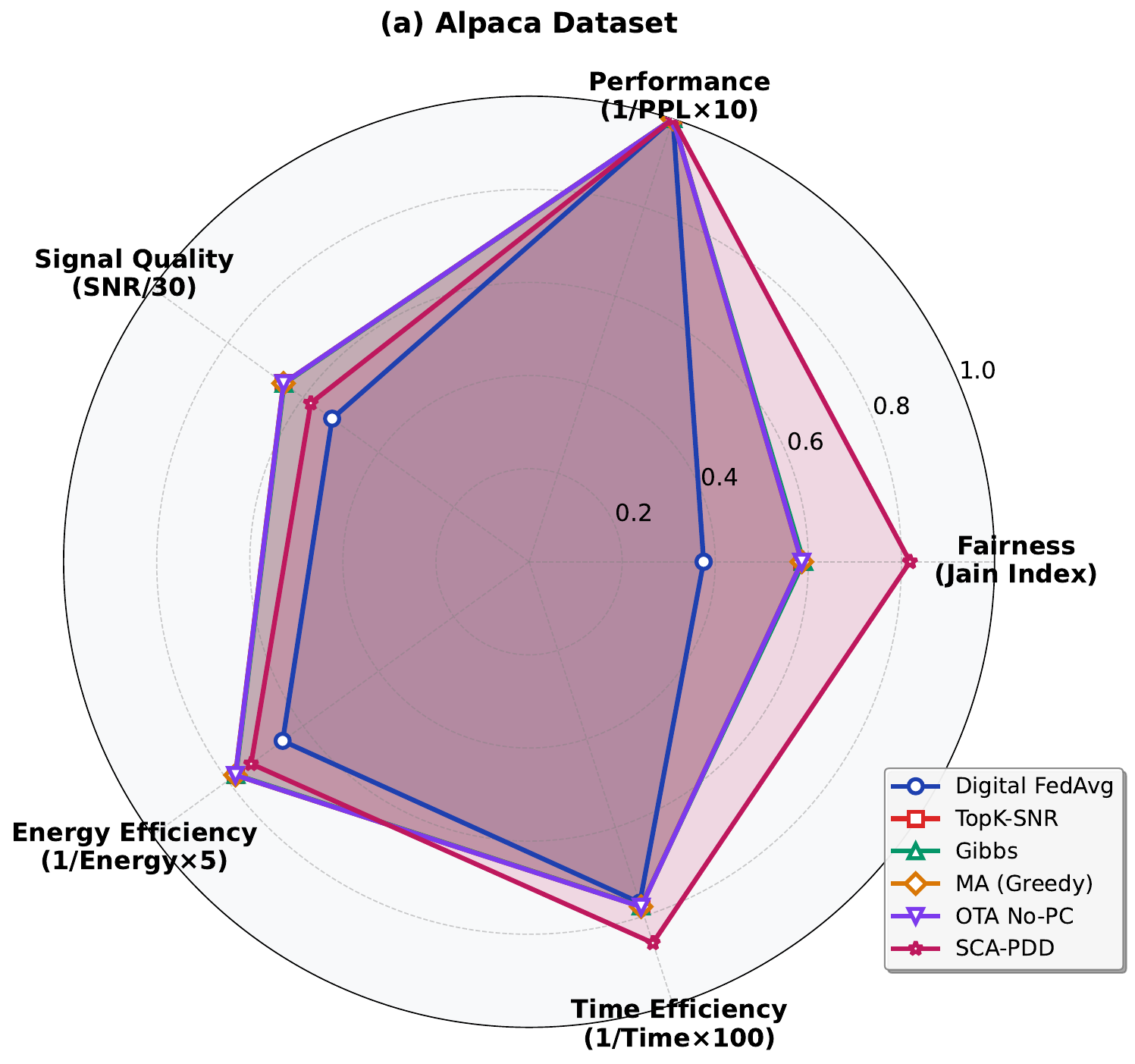}
  \caption{Performance comparison of different learning methods on Alpaca dataset.}
  \label{fig:snr-alp}
\end{figure}

Average SNR summarizes radio conditions but does not by itself predict learning outcomes under analog aggregation (Fig.~\ref{fig:snr-alp}). 
At $k{=}1$, the highest‑SNR baseline (OTA No‑PC) reports 24.2\,dB, yet its perplexity is worse than SCA–PDD, which operates at 21.7\,dB. 
At $k{=}2$, the contrast is stronger: 28.5\,dB (OTA No‑PC) versus 19.7\,dB (SCA–PDD), while perplexity still favors SCA–PDD (3.03 vs.\ 3.53). 
Even at $k{=}4$, SCA–PDD achieves competitive perplexity at a lower average SNR (16.3 dB) than TopK/Gibbs (18.7 dB).

\subsection{Dolly Experiments}
\label{sec:dolly-num}

We evaluate on \textsc{Dolly}\,15k ~\cite{databricks2023dolly} under the same communication and training budgets as in Sec.~\ref{sec:alpaca}. Ten clients each hold a disjoint shard. We vary the per–round concurrency $k\in\{1,2,4,8\}$ and report all metrics at the R30 horizon.

\subsubsection{Fairness}
Fairness is most difficult when few clients superpose per round as shown in Fig.~\ref{fig:jain} (e)-(h). In this regime, SCA-PDD substantially widens participation: at $k{=}1$ the Jain index is $0.789$ versus $0.277$ for the strongest comparator (Gibbs), an absolute gain of $0.512$ (relative $184.8\%$). At $k{=}2$ the index reaches $0.904$ versus $0.441$ (gain $0.463$, $105.0\%$). As concurrency increases, all methods approach the fairness ceiling simply because more clients are active each round, but SCA–PDD remains competitive: $0.970$ versus $0.774$ in $k{=}4$ (gain $0.196$, $25.3\%$) and $0.985$ in $k{=}8$, within $\approx 0.2\%$ of the highest value ($0.987$ for MA~Greedy). These results indicate that geometry–aware scheduling counters the tendency of SNR–centric heuristics to concentrate airtime on a handful of strong links, particularly where fairness is hardest (small $k$).

\subsubsection{Inequality}
The inequality view mirrors the fairness gains and clarifies tail behavior in Fig.~\ref{fig:gini} (e)-(h). At $k{=}1$, Gini falls to $0.286$ from $0.666$ (Gibbs), a reduction of $0.380$ ($57.1\%$). At $k{=}2$ it drops to $0.176$ from $0.537$ ($0.361$, $67.2\%$). At $k{=}4$, SCA–PDD attains $0.100$ versus $0.286$ ($0.186$, $65.0\%$). At $k{=}8$, the methods converge towards uniformly high participation; SCA–PDD records $0.069$ while the lowest baseline is $0.064$ (MA~Greedy). In effect, coupling client rotation with beamforming and MA geometry “compresses the tail”: weaker links appear more frequently without destabilizing analog aggregation.

\subsubsection{Model quality}

Quality improvements at a fixed round budget are most pronounced at small–moderate concurrency (Fig.~\ref{fig:ppl}). At $k{=}1$, perplexity is $4.62$ compared with $10.12$ for the strongest comparator, an absolute reduction of $5.50$ ($54.3\%$). At $k{=}2$, PPL is $5.10$ versus $5.85$ (MA~Greedy), a reduction of $0.75$ ($12.8\%$). For $k{=}4$, SCA–PDD is within $0.15$ of the best value ($5.29$ vs.\ $5.14$ for Gibbs, $2.9\%$ difference), while simultaneously delivering substantially higher fairness (Jain $0.970$ vs.\ $0.764$–$0.774$) and lower inequality (Gini $0.100$ vs.\ $0.286$–$0.297$). At $k=8$, Digital~FedAvg achieves the lowest PPL ($4.75$), reflecting the absence of analog mismatch; SCA–PDD reports $5.16$ with nearly saturated fairness. In general, the data suggest a practical operating region in $k{=}1$–$2$, where minimizing the analog mismatch translates more effectively into better fixed–budget model quality.

\subsubsection{Link quality and efficiency}

\begin{figure}[t]
  \centering
  \includegraphics[width=0.7\columnwidth]{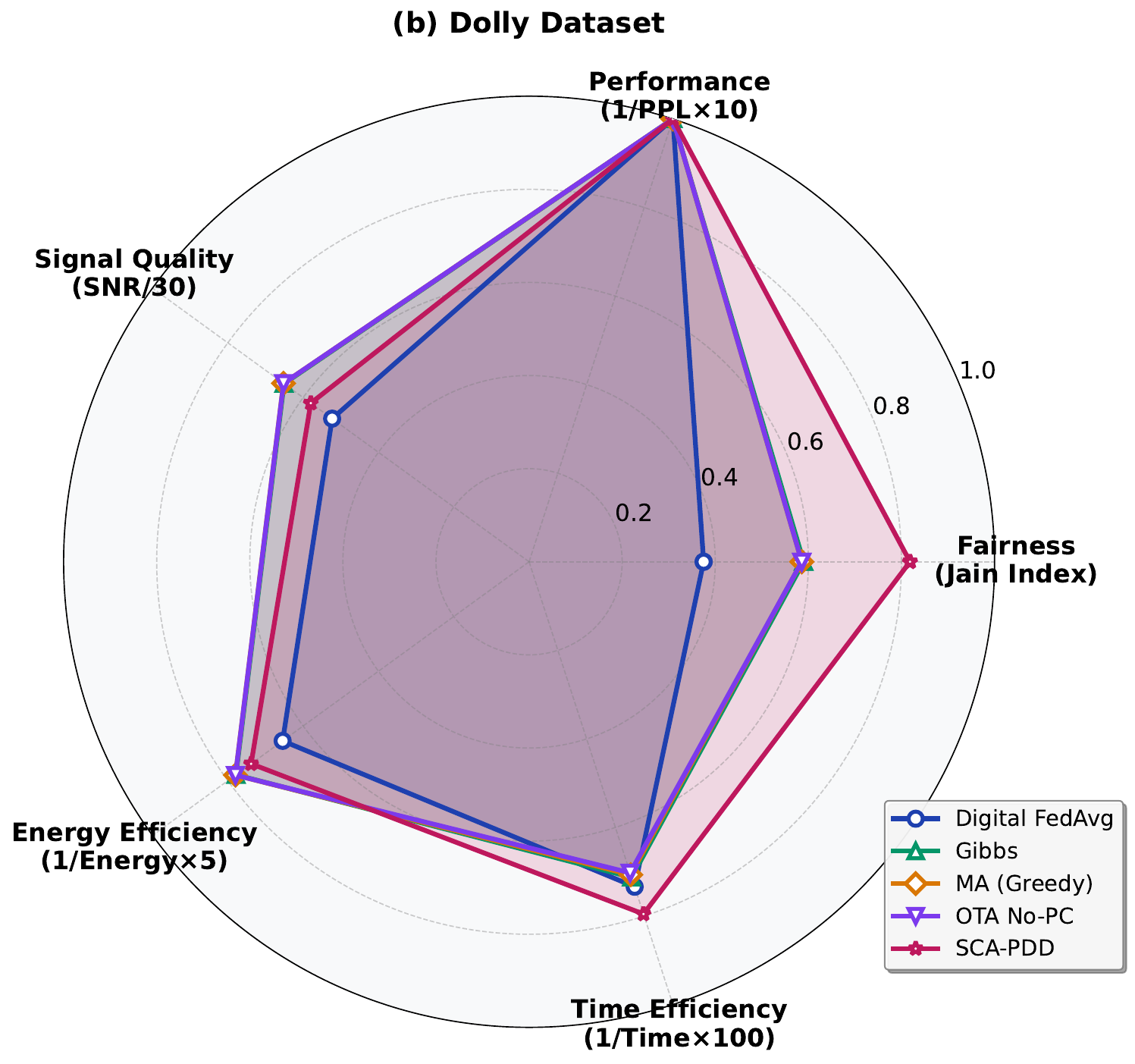}
  \caption{Performance comparison of different learning methods on Dolly dataset.}
  \label{fig:snr-dolly}
\end{figure}

Average post–combining SNR is informative about RF conditions but does not by itself predict learning under analog aggregation (Fig.~\ref{fig:snr-dolly}). At $k{=}1$, SCA–PDD operates at $21.8$ dB while the highest–SNR comparator reaches $24.3$\,dB, yet SCA–PDD more than halves the perplexity. At $k{=}2$, the SNR gap widens (19.7 dB vs. 28.6 dB), with SCA–PDD still attaining a lower PPL. Even at $k{=}4$, where SCA–PDD’s SNR is slightly higher (16.3 dB vs. 15.6–15.7 dB), the best baseline keeps a small PPL edge. This behavior is consistent with the analysis in Sec.~\ref{sec:Convergence}: the effective learning signal depends on both noise and the \emph{vector mismatch} between the received analog sum and the desired weighted gradient. Maximizing SNR alone is therefore not a reliable proxy for downstream quality; controlling mismatch through geometry and beamforming is decisive at small to moderate $k$.

\subsection{Discussion}

Across datasets and concurrency levels, two consistent observations emerge. First, client rotation coupled with geometry‑aware alignment yields both higher fairness and lower perplexity within the same round budget, as seen by the separation of the Jain/Gini curves and the PPL bars at R30. Second, SNR alone is not predictive of model quality in the analog regime: the radar plots show the proposed method can outperform in PPL while not maximizing SNR, underscoring that alignment of the received analog sum with the intended weighted gradient is the key optimization target.

\section{Conclusion}
\label{sec:conclusion}
A movable‑antenna framework was integrated with over‑the‑air aggregation for federated fine‑tuning of large language models. The design jointly optimizes client participation, local computation and batching, MA element placement, and analog beamforming under latency and energy constraints, solved with a hybrid SCA–PDD routine. The analysis accounts for OTA‑induced distortion and distribution shift between pre‑training and fine‑tuning. Experiments show improvements in model quality and participation fairness across multiple uplink concurrencies, together with favorable latency–energy trade‑offs. Aligning the received analog sum with the intended weighted gradient is more decisive than maximizing post‑combining SNR when only a few clients transmit concurrently.

\bibliographystyle{ieeetr}
\bibliography{related}

\newpage
\appendix

\subsection{Complexity Analysis}
\label{sec:Complexity}
During each SCA--PDD iteration, the Block A SCA subproblem is convex and can be solved in polynomial time (e.g.,\ \(\mathcal{O}(p^{\alpha})\) for a problem dimension \(p\)).  The PDD often uses coordinate or projected gradient methods to handle discrete or geometric constraints.

Because each \(\bigl(M,N\bigr)\) is enumerated, the total complexity is roughly
\begin{equation}
  \label{eq:sca-pdd-complexity}
  \mathcal{O}\Bigl(
    |\mathcal{M}|\;\times\;|\mathcal{N}|\;\times\;I_{\text{SCA}}\;\times\;p^{\alpha}
  \Bigr),
\end{equation}
where \(\mathcal{M}\) and \(\mathcal{N}\) are finite sets of candidate integer values for \(M\) and \(N\), \(I_{\text{SCA}}\) is the typical number of SCA--PDD iterations and \(\alpha\in [2,3]\) depends on the solver.  In practice, moderate \((M,N)\) ranges and a small number of SCA--PDD iterations often suffice, making the approach tractable.

\subsection{Convergence Analysis}
\label{sec:Convergence}

We analyze a two-phase LLM training procedure (Phases\~I--II) under a potential distribution shift, with OTA gradient aggregation in the fine-tuning phase. Concretely, our aim is to characterize how the final fine-tuning loss depends on several key parameters: (i) the number of pre-training rounds $M$ and fine-tuning rounds $N$, (ii) the variance bounds $\alpha^2$ and $\hat{\alpha}^2$ together with the smoothness constants $\rho$ and $\hat{\rho}$~\cite{stich2019localsgdconvergesfast}, (iii) the Wasserstein distance $\mathrm{W}(P_{\mathrm{pre}}, P_{\mathrm{fine}})$ governing the distribution shift~\cite{DBLP:journals/corr/abs-2007-01434,shen2018wassersteindistanceguidedrepresentation}, and (iv) the OTA noise variance $\sigma^2$, which is affected by MA placements $\{\mathbf{x}_{n,i}\}$ and beamforming vectors $\{\mathbf{q}^{(n)}\}$.

To facilitate our analysis, we first assume that $L_{\mathrm{pre}}$ is $\rho$-smooth, meaning that for all $\mathbf{w}_1,\mathbf{w}_2$,
\begin{equation}
\label{eq:pre-smooth}
\begin{split}
   L_{\mathrm{pre}}(\mathbf{w}_1)
   \;\le\;
   L_{\mathrm{pre}}(\mathbf{w}_2)
   \;+\;
   \nabla L_{\mathrm{pre}}(\mathbf{w}_2)^{T}\bigl(\mathbf{w}_1 - \mathbf{w}_2\bigr)
   \\
   +\; \tfrac{\rho}{2}\|\mathbf{w}_1 - \mathbf{w}_2\|^2,
\end{split}
\end{equation}
and that $L_{\mathrm{fine}}$ is $\hat{\rho}$-smooth, i.e., for all $\mathbf{w}_1,\mathbf{w}_2$,
\begin{equation}
\label{eq:fine-smooth}
\begin{split}
   L_{\mathrm{fine}}(\mathbf{w}_1)
   \;\le\;
   L_{\mathrm{fine}}(\mathbf{w}_2)
   \;+\;
   \nabla L_{\mathrm{fine}}(\mathbf{w}_2)^{T}\bigl(\mathbf{w}_1 - \mathbf{w}_2\bigr)
   \\
   +\; \tfrac{\hat{\rho}}{2}\|\mathbf{w}_1 - \mathbf{w}_2\|^2.
\end{split}
\end{equation}

In addition, each training phase has bounded gradient variance: $\mathbb{E}\bigl[\|\nabla \tilde{L}_{\mathrm{pre}}(\mathbf{w}) - \nabla L_{\mathrm{pre}}(\mathbf{w})\|^2\bigr] \le \alpha^2$ during Phase~~I, and $\mathbb{E}\bigl[\|\nabla \tilde{L}_{\mathrm{fine}}(\mathbf{w}) - \nabla L_{\mathrm{fine}}(\mathbf{w})\|^2\bigr] \le \hat{\alpha}^2$ during Phase~~II. To capture the distribution shift between the pretraining distribution $P_{\mathrm{pre}}$ and the fine-tuning distribution $P_{\mathrm{fine}}$, we use the Wasserstein distance $\mathrm{W}(P_{\mathrm{pre}}, P_{\mathrm{fine}})$. Specifically, if $\ell(\mathbf{w}, \mathbf{d})$ is $\rho_{\mathrm{dist}}$-Lipschitz in $\mathbf{d}$, then
\begin{equation}
\label{eq:wasserstein-shift}
   \bigl|\,L_{\mathrm{fine}}(\mathbf{w}) - L_{\mathrm{pre}}(\mathbf{w})\bigr|
   \;\le\;
   \rho_{\mathrm{dist}}\,\mathrm{W}(P_{\mathrm{pre}}, P_{\mathrm{fine}}).
\end{equation}

Finally, in the fine-tuning phase (Phase\~II), the server observes aggregated gradients
$\widehat{\mathbf{G}}^{(n)} = \mathbf{G}^{(n)} + \boldsymbol{\varepsilon}^{(n)}$,
where $\boldsymbol{\varepsilon}^{(n)}$ denotes the OTA noise, satisfying $\mathbb{E}\bigl[\|\boldsymbol{\varepsilon}^{(n)}\|^2\bigr] \le \sigma^2$. This noise level $\sigma^2$ can be mitigated by optimizing position of MA and beamforming design.

\subsubsection{Phase I: Pre-training Convergence}
We first perform $M$ rounds of mini-batch SGD on $L_{\mathrm{pre}}$. Specifically, the update rule is
\begin{subequations}
\begin{align}
  \label{eq:phase1-update}
  \mathbf{w}^{(m+1)}
  &= 
  \mathbf{w}^{(m)}
  - 
  \gamma \,\nabla \tilde{L}_{\mathrm{pre}}\bigl(\mathbf{w}^{(m)}; \zeta^{(m)}\bigr)
  \\
  \label{eq:phase1-index}
  m &= 0,\dots,M-1
\end{align}
\end{subequations}
where $\zeta^{(m)}$ is a mini-batch drawn from the pre-training distribution $P_{\mathrm{pre}}$, and $\gamma$ is the step size. Under the assumption that $L_{\mathrm{pre}}$ is $\rho$-smooth and that the stochastic gradient has variance at most $\alpha^2$, standard SGD analysis implies the following bound:
\begin{equation}
  \label{eq:PreTrainStandard}
  \begin{split}
    \frac{1}{M}\sum_{m=0}^{M-1}
      \mathbb{E}\bigl[\|\nabla L_{\mathrm{pre}}(\mathbf{w}^{(m)})\|^2\bigr]
    \;\le\;
    \frac{2\bigl(L_{\mathrm{pre}}(\mathbf{w}^{(0)}) - L_{\mathrm{pre}}^{*}\bigr)}{\gamma\,M}
    \\
    +\; \rho\,\gamma\,\alpha^2.
  \end{split}
\end{equation}
From this, one can equivalently derive an upper bound on
$\mathbb{E}[L_{\mathrm{pre}}(\mathbf{w}^{(M)})]$ that depends on $M, \gamma, \rho,$ and $\alpha^2.$ We denote this dependency by $\Delta_{\mathrm{pre}}(M,\gamma,\alpha^2,\rho)$.

\subsubsection{Distribution Shift Bound}
Upon completion of Phase~I, the parameter vector is $\mathbf{w}^{(M)}$. To account for the distribution shift between $P_{\mathrm{pre}}$ and $P_{\mathrm{fine}}$, we assume that there is a bound on $|\,L_{\mathrm{fine}}(\mathbf{w}) - L_{\mathrm{pre}}(\mathbf{w})|$ proportional to the Wasserstein distance $\mathrm{W}(P_{\mathrm{pre}},P_{\mathrm{fine}})$. Concretely,
\begin{equation}
  \label{eq:fine-from-pre-M}
  L_{\mathrm{fine}}\bigl(\mathbf{w}^{(M)}\bigr)
  \;\le\;
  L_{\mathrm{pre}}\bigl(\mathbf{w}^{(M)}\bigr)
  \;+\;
  \rho_{\mathrm{dist}} \,
  \mathrm{W}(P_{\mathrm{pre}},P_{\mathrm{fine}}).
\end{equation}
Taking expectation over the randomness of the SGD updates, we obtain
\begin{equation}
  \label{eq:AfterPhaseI}
  \mathbb{E}\bigl[L_{\mathrm{fine}}(\mathbf{w}^{(M)})\bigr]
  \;\le\;
  \mathbb{E}\bigl[L_{\mathrm{pre}}(\mathbf{w}^{(M)})\bigr]
  \;+\;
  \rho_{\mathrm{dist}}\,\mathrm{W}.
\end{equation}
Finally, by using our Phase~I bound on 
$\mathbb{E}[L_{\mathrm{pre}}(\mathbf{w}^{(M)})]$, 
we obtain an upper bound on 
$\mathbb{E}[L_{\mathrm{fine}}(\mathbf{w}^{(M)})]$ of the form
\begin{equation}
\label{eq:AfterPhaseI-final}
\begin{split}
  \mathbb{E}\bigl[L_{\mathrm{fine}}(\mathbf{w}^{(M)})\bigr]
  \;\le\;
  L_{\mathrm{pre}}\bigl(\mathbf{w}^{(0)}\bigr)
  \;+\;
  \Delta_{\mathrm{pre}}\bigl(M,\gamma,\alpha^2,\rho\bigr)
  \\
  +\; \rho_{\mathrm{dist}}\,\mathrm{W}.
\end{split}
\end{equation}

\subsubsection{Phase II: Fine-Tuning with OTA Noise}
Next, in Phase~II we fine-tune the model for $N$ additional rounds on $L_{\mathrm{fine}}$, but with OTA gradient aggregation subject to additive noise. Concretely, each round $n=0,\dots,N-1$ is
\begin{equation}
  \label{eq:phase2-update}
  \mathbf{w}^{(M+n+1)}
  \;=\;
  \mathbf{w}^{(M+n)}
  \;-\;
  \hat{\gamma}\,\widehat{\mathbf{G}}^{(n)},
\end{equation}
where $\widehat{\mathbf{G}}^{(n)} = \mathbf{G}^{(n)} + \boldsymbol{\varepsilon}^{(n)}$. Here, $\mathbf{G}^{(n)}$ denotes the aggregated user gradients (variance at most $\hat{\alpha}^2$), and $\boldsymbol{\varepsilon}^{(n)}$ represents OTA noise with variance at most $\sigma^2$~\cite{8952884}. Thus, the total variance in the fine-tuning phase is $\hat{\alpha}^2 + \sigma^2$. 

Under the assumption that $L_{\mathrm{fine}}$ is $\hat{\rho}$-smooth, a standard SGD argument yields~\cite{stich2019localsgdconvergesfast}:
\begin{align}
\frac{1}{N} \sum_{n=0}^{N-1}
    \mathbb{E}\bigl[\|\nabla L_{\mathrm{fine}}(\mathbf{w}^{(M+n)})\|^2\bigr]
&\;\le\;
  \frac{2\bigl(L_{\mathrm{fine}}(\mathbf{w}^{(M)}) - L_{\mathrm{fine}}^{*}\bigr)}{\hat{\gamma}\,N}
  \nonumber\\
&\quad
  +\;\hat{\rho}\,\hat{\gamma}\,\bigl(\hat{\alpha}^2 + \sigma^2\bigr).\label{eq:phase2-grad-bound}
\end{align}
By rearranging the above bound, we obtain a direct upper bound on $\mathbb{E}[L_{\mathrm{fine}}(\mathbf{w}^{(M+N)})]$, which we denote by
$\Delta_{\mathrm{fine}}(N,\hat{\gamma},\hat{\alpha}^2+\sigma^2,\hat{\rho})$.

In summary, the distribution-shift result \eqref{eq:AfterPhaseI}, and the Phase~II analysis leads to the final convergence guarantee
\begin{align}
\mathbb{E}\bigl[L_{\mathrm{fine}}(\mathbf{w}^{(M+N)})\bigr]
&\;\le\;
\underbrace{L_{\mathrm{pre}}(\mathbf{w}^{(0)})}_{\text{initial}}
\;+\;
\underbrace{\Delta_{\mathrm{pre}}(M,\gamma,\alpha^2,\rho)}_{\text{Phase I}}
\nonumber\\[6pt]
&\quad
+\;\underbrace{\rho_{\mathrm{dist}}\;\mathrm{W}}_{\text{shift}}
\;+\;
\underbrace{\Delta_{\mathrm{fine}}(N,\hat{\gamma},\hat{\alpha}^2 + \sigma^2,\hat{\rho})}_{\text{Phase II}}.\label{eq:final-convergence}
\end{align}
Intuitively, the total expected fine-tuning loss is bounded by: (i) the initial loss $L_{\mathrm{pre}}(\mathbf{w}^{(0)})$, (ii) the overhead of pre-training convergence $\Delta_{\mathrm{pre}}$, (iii) the penalty for distribution shift $\rho_{\mathrm{dist}}\,\mathrm{W}$~\cite{DBLP:journals/corr/abs-2007-01434,shen2018wassersteindistanceguidedrepresentation}, and (iv) the overhead of fine-tuning convergence $\Delta_{\mathrm{fine}}$, which grows with both $\hat{\alpha}^2$ and the OTA noise variance $\sigma^2$. This result cleanly reveals how each factor influences the final performance.

\end{document}